\documentclass[10pt,journal,final,twoside,letterpaper]{%
  ./etc/IEEEtran}

\usepackage{%
  ./etc/paper_header}
\usepackage{%
  algorithm,algorithmic}

\newcommand{\mbb}{\mathbb}

\allowdisplaybreaks

\begin{document}


\title{%
  An Efficient Polyphase Filter Based Resampling Method for Unifying the PRFs in SAR Data}

\author{%
  Yoangel~Torres,  
  Kamal~Premaratne,~\IEEEmembership{Senior~Member,~IEEE,} 
  Falk~Amelung, 
  and~Shimon~Wdowinski
  \thanks{%
    Manuscript submitted in May 2017.}
  \thanks{%
    YT is at the Northrup Grumman Corporation, Melbourne, Florida; this work was conducted while he was at the University of Miami (UM), Coral Gables, Florida (e-mail: y.torres1@umiami.edu). KP is with the Department of Electrical and Computer Engineering, UM. FA is with the Department of Marine Geosciences, Rosenstiel School of Marine and Atmospheric Sciences (RSMAS), UM (e-mail: \{kamal, f.amelung\}@miami.edu). SW is with the Department of Earth \& Environment, Florida International University; this work was conducted while he was with RSMAS, UM (e-mail: shimon.wdowinski@fiu.edu).}
  \thanks{%
  This work is based on research supported by the UM College of Engineering/RSMAS Collaborative Research Initiative in Environmental Sensing, U.S. Office of Naval Research (ONR) via grant \#N00014-10-1-0140, and the U.S. National Science Foundation (NSF) via grant \#1343430.}}
  \markboth{%
  IEEE Transactions on Geoscience and Remote Sensing \fbox{In Review}{}}{%
  Torres, \MakeLowercase{\ti{et al.}}: Polyphase Filter Based Resampling Method for Unifying PRFs}

\IEEEpubid{0000--0000/00\$00.00~\copyright~2017 IEEE}
\maketitle


\begin{abstract} 
Variable and higher pulse repetition frequencies (PRFs) are increasingly being used to meet the stricter requirements and complexities of current airborne and spaceborne synthetic aperture radar (SAR) systems associated with higher resolution and wider area products. \emph{POLYPHASE,} the proposed resampling scheme, downsamples and unifies variable PRFs within a single look complex (SLC) SAR acquisition and across a repeat pass sequence of acquisitions down to an effective lower PRF. A sparsity condition of the received SAR data ensures that the uniformly resampled data approximates the spectral properties of a decimated densely sampled version of the received SAR data. While experiments conducted with both synthetically generated and real airborne SAR data show that POLYPHASE retains comparable performance to the state-of-the-art BLUI scheme in image quality, a  polyphase filter-based implementation of POLYPHASE offers significant computational savings for arbitrary (not necessarily periodic) input PRF variations, thus allowing fully on-board, in-place, and real-time implementation. 
\end{abstract}




\section{Introduction}
\label{sec:Introduction}

 
Synthetic aperture radar (SAR) and interferometric SAR (InSAR) find application in geophysical and environmental remote sensing applications \cite{Moreira2013GRSM}. However, as sensor technology advances, the complexity of how the data are collected also increases. Variation in pulse repetition frequency (PRF) is one of these complexities introduced by modern sensors. 

Methods that make use of displaced phase centers (DPCs) are available to recover the unambiguous Doppler spectrum from non-uniform spatial sampling of the synthetic aperture. A simple two-point interpolation and multichannel reconstruction scheme appears in \cite{krieger2004sar, gebert2005, gebert2010multichannel}. An innovative frequency domain algorithm that enables unambiguous recovery of the Doppler spectrum in the case of a single channel appears in \cite{Jiang2012}. A computationally efficient time domain scheme which handles single channel non-uniform oversampled SAR data generated from a platform accelerating along-track by resampling the data in the slow-time domain is in \cite{Goldman2010RC}. However, this method assumes a constant PRF as the platform moves along track. The spatial non-uniformity is solely due to small changes in velocity arising from uncontrolled acceleration of the platform, which is typically much smaller than what could be generated from variable PRFs. 

\IEEEpubidadjcol

\emph{POLYPHASE,} the new scheme we propose, tackles non-uniformity along-track within a single look complex (SLC) single channel or post-beamformed SAR collection arising from different PRFs (or from arbitrary sampling). It takes in demodulated SAR data for different acquisitions, which are collected and oversampled at variable PRFs, and delivers resampled data at a lower, constant PRF within each acquisition, and uniformly sampled in the spatial frequency domain ($k$-space) \cite{Walker1980ToAE, Soumeck1994}. A new polyphase filter-based implementation allows digital filtering at the lowest possible rate, viz., the effective output PRF rate. The result is a computationally efficient fully on-board algorithm enabling in-place and real-time processing which avoids up/down-link data transfers and bottlenecks. The POLYPHSE method approximately reconstructs the collected data on a uniformly spaced grid along the synthetic aperture, while preserving the resolution and Nyquist constraint within the cross-range extent of interest. 

We use the spectral properties of the SAR data to justify, and real SAR data to verify, the proposed POLYPHASE scheme. The  \emph{best linear unbiased interpolation (BLUI)} scheme in \cite{Villano2014ToGRS} also uses spectral properties to interpolate between non-uniformly oversampled SAR data. However, the type of antenna and the type of noise present may render BLUI sub-optimal because of the need to estimate the SNR and numerically evaluate the autocorrelation function of the SAR signal. BLUI also imposes a minimum aperture length (i.e., antenna size) and a maximum platform velocity so that enough number of samples contribute to the interpolation. POLYPHASE simply mandates a lower bound on the sparseness of the received SAR data relative to the output grid. It ensures delivery of uniformly resampled data which approximates the spectral properties of a decimated version of a `hidden' densely sampled SAR data sequence which can be considered to have generated the non-uniformly sampled input SAR data. 

Moreover, POLYPHASE works with arbitrary input PRF variations. Thus, it is applicable in more general scenarios (e.g., to compensate for flight path deviation, imaging while in turn, and uncontrolled platform acceleration/deceleration effects \cite{Goldman2010RC}). In contrast, BLUI imposes certain periodicity constraints on the input PRF variation to realize its computational savings. In other words, BLUI caters well to the staggered SAR scenario \cite{Villano2015APSAR}, but not for arbitrary PRF variations.

Section~\ref{sec:Tech} provides a summary of the technical background. Section~\ref{sec:BasicIdea} provides the main idea, and Sections~ \ref{sec:OutputGrid} and \ref{sec:filter} more details, of the POLYPHASE scheme. Sections~\ref{sec:Results} and \ref{sec:Discussion} illustrate the application of POLYPHASE and the ensuing results. Section~\ref{sec:Conclusion} provides concluding remarks. 

\begin{table*}[!t]
  \centering
  \caption{Notation}
  \vv
  \scriptsize
  \renewcommand{\arraystretch}{1.1}\addtolength{\tabcolsep}{-2pt}
  \begin{tabular}{lp{5.5in}} 
    \hline
    \hfil\tb{Notation}
      & \hfil \tb{Description} \\
    \hline\hline 
    $\mbb{C}$, $\mbb{R}$, $\mbb{N}$, $\mbb{N}_+$
      & Complex numbers, real numbers, integers, and non-negative integers, respectively. \\ 
    $\mbb{C}^{M\times N}$, $\mbb{R}^{M\times N}$
      & $(M\times N)$-sized matrices with complex- and real-valued entries, respectively. \\ 
    $(x)_L$
      & For $x\in\mbb{R}$, remainder of the dividend $x\in\mbb{R}$ when $L\in\mbb{N}_+$ is the divisor. \\
    $\lfloor{x}\rfloor$, $\lceil{x}\rceil$
      & For $x\in\mbb{R}$, integer value not more than $x$ and integer value not less than $x$, respectively. \\
    $\delta(n)$, $\delta_D(\omega)$
      & Kronecker delta function for $n\in\mbb{N}_+$ and the Dirac delta function for $\omega\in\mbb{R}$, respectively. \\
    \hline
    $c$, $V_p$
      & Speed of light (m/s) and radar platform velocity (m/s), respectively. \\
    $\lambda_{\min}$, $\lambda_c$, $\lambda_{\max}$
      & Minimum, center, and maximum radar transmit wavelength (m), respectively. \\
    $B_{chirp}$, $PBW$
      & Transmit chirp bandwidth (Hz) and processing Doppler bandwidth (Hz),  respectively. \\
    $D$, $D_{pad}$
      & Synthetic aperture length (m) and padded synthetic aperture length (m), respectively. \\
    $R$, $R_0$
      & Slant range between the radar and an arbitrary point on scene (m) and the center of scene (m), respectively. \\
    $X_{\tx{in}}$, $\beta$, $\Delta\beta$
      & Cross-rage extent (m), squint angle (rad), and beam sweep angle (rad), respectively. \\  
    $\theta_{\min}$, $\theta_{\max}$
      & Minimum and maximum look angles, respectively, seen by the synthetic aperture (rad). \\   
    \hline
    $t$, $\ol{f}$, $\nu$
      & Temporal dimension (s), temporal frequency (Hz), and its rotational counterpart (rad/s) respectively. $\nu=2\pi\ol{f}$. \\
    $u$, $\ol{h}$, $\omega$
      & Spatial dimension (m), spatial frequency (1/m), and its rotational counterpart (rad/m), respectively. $\omega=2\pi\ol{h}$. \\
    $\omega_r$, $\omega_{cr}$
      & Wavenumber in range (rad/m) and cross-range (azimuth) (rad/m), respectively. $\omega_r=\nu_r/c$, $\omega_{cr}=\nu_{cr}/V_p$. \\  
    $\delta_r$, $\delta_{cr}$
      & Resolution in range and azimuth, respectively. \\
    $K_r$, $K_{cr}$
      & Broadening factors for pulse weighting in range and aperture weighting in azimuth (or cross-range), respectively. \\
    \hline
    $(x, y, z)$
      & (cross-range, range, height) coordinates of the Cartesian coordinate system (m); scene center is at $(0, 0, 0)$. \\
    $(X_{\tx{out}}, Y_{\tx{out}})$, $(X_n, Y_n, Z_n)$
      & (Azimuth, range) dimensions of the imaged scene (m), 
        Coordinates of the $n$-th point scatterer within the imaged scene (m). \\
    $u,\,u_k,\,u_{\min},\,u_{\tx{mid}},\,u_{\max}$
      & Continuous-time $x$-coordinate, discrete-time $x$-coordinate, minimum, mid, and maximum $x$-coordinates, respectively, of the synthetic aperture (m). \\
    $\Delta u_{\tx{in}}$, $\Delta u_{\tx{out}}$
      & Along-track sample and resample spacings at boresight (m), respectively. \\    
    \hline
    $N_{FFT}$
      & Fast Fourier transform (FFT) size used for spatial, or azimuth compression, FFT computation (after zero-padding). \\
    $N_{X\!AC}$, $N_D$, $p_d$  
      & Number of discrete points needed with and without boundary regions to represent the synthetic aperture length without aliasing and spatial frequency spectrum oversampling factor, respectively. $p_d=(N_{FFT}/N_D)\,K_{cr}$. \\   
    $PRF_{\tx{in}}$, $PRF_{\tx{out}}$
      & Input and output pulse repetition frequencies, respectively (1/m). \\
    $PRI_{\tx{in}}$, $PRI_{\tx{out}}$
      & Input and output pulse repetition intervals, respectively (s). $PRF_{\tx{in}}=1/PRF_{\tx{in}}$, $PRF_{\tx{out}}=1/PRF_{\tx{out}}$. \\
    $\gamma$
      & Normalized processing bandwidth. $\gamma=PBW/PRF_{\tx{out}}$. \\
    \hline
    \end{tabular}
  \label{tab:notation}
\end{table*}


\section{Technical Background}
\label{sec:Tech}


We use a Cartesian coordinate system with the origin at the scene center, $x$-axis along-track and parallel to the SAR platform velocity vector $V_p$ (m/s), and $y$-axis along boresight; $z$-axis denotes altitude \cite{Wehner1987, Carrara1995, Jakowatz1996, Soumeck1999}. Table~\ref{tab:notation} shows the notation. 

\bi{Variable PRFs.} 
While conventional radar system operation relies on a constant PRF, technological advances now allow for newer radar modes of operation, e.g., the high-resolution, wide-swath imaging in multi-channel SAR allows a shorter revisit time for frequent global mapping. In wide-swath imaging, the antenna length limitation which can restrict the achievable swath width is overcome by a technique based on a single azimuth channel with the system operating with a continuously varied PRF \cite{Gebert2010EUSAR}. This allows arbitrary wide swaths and distributes the discrete blind ranges according to the applied PRF span of values. In the end, continuous coverage is achieved at the cost of partial blockage (i.e., loss of some pulses for every target). This PRF variation manifests itself as non-uniform sampling of the slow-time domain along the synthetic aperture \cite{Gebert2010EUSAR}, thus requiring additional processing, e.g., interpolation schemes to resample the signal to a regular azimuth grid \cite{Gebert2010EUSAR}.  

The advantages offered by high-resolution, ultra-wide swath SAR imaging is also exploited in multiple elevation beam (MEB) SAR based on variable PRF \cite{Yadong2013IRC}, which employs digital beamforming with a reflector antenna to improve SNR and suppress range ambiguities. It also employs linear variation of the pulse repetition interval (PRI) to overcome the blind range problem of conventional MEB SAR \cite{Luo2014}. 

These new techniques of high-resolution, wide-swath imaging modes \cite{Gebert2010EUSAR, Yadong2013IRC, Luo2014} come at the cost of non-uniform sampling of the slow-time along-track Doppler phase. Spatial discrete Fourier transform (DFT) processing of such non-uniformly spaced data can introduce undesirable artifacts (e.g., smearing, defocusing, and echoing) into the final image (see Fig.~\ref{fig:Dataset12}).


\section{Proposed POLYPHASE Resampling Scheme}
\label{sec:BasicIdea}


While the POLYPHASE scheme applies to both strip and spotlight mode SAR, concentrating on the latter, the underlying complex-valued signal of interest $s(\cdot,\cdot)$ is taken as \cite{Soumeck1999}
\begin{equation}
  s(\nu_r,u) 
    =S_p(\nu_r)
     \sum_N
     \sigma_n 
     e^{-j2\omega_r D_s(n)}.
  \label{eq:SARsignal}
\end{equation}
Here, $\nu_r$ is the rotational frequency (rad/s) (in range), $\omega_r$ is the corresponding wave number (rad/m), $S_p(\nu_r)$ is the discrete-time (DT) Fourier transform (FT) of the transmitted signal $p(t)$ (in range), $\sigma_n$ is the reflectivity of the $n$-th scatterer in the scene, $D_s(n)=\sqrt{(X_n-u)^2+(Y_n-Y_c)^2+(Z_n-Z_c)^2}$, $(X_n, Y_n, Z_n)$ are its coordinates, $(u, Y_c, Z_c)$ are the instantaneous radar coordinates, and $u\in [u_{\min}, u_{\max}]$ is the along-track platform position (m). The ground range and altitude coordinates $(Y_c, Z_c)$ are taken to be constants.  


\subsection{Uniformly Sampled Radar Signal}
\label{subsec:sn}

\begin{figure}[!t]
  \centering
  \includegraphics[width=0.48\textwidth]{%
    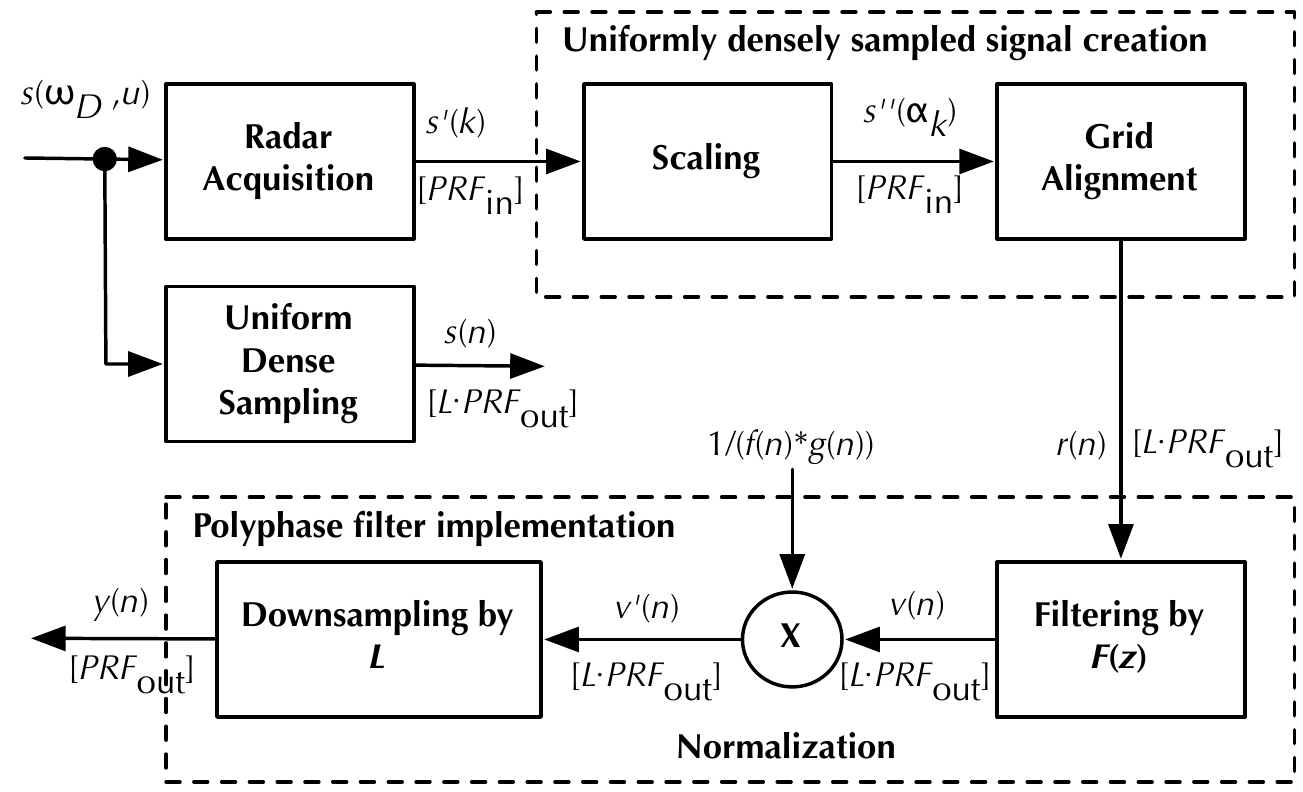}
  \vv
  \caption{\scriptsize{Block diagram of the proposed POLYPHASE scheme. The sampling rate at each stage is shown within square brackets (e.g., $[PRF_{\tx{in}}]$).}}
  \label{Fig:UnifiedBlockDiagram}
\end{figure}

Let us uniformly sample the radar signal $s(\nu_r, u),\,u\in[u_{\min}, u_{\max}]$, where $u_{\min}\leq u_0\leq\cdots\leq u_{K-1}\leq u_{\max}$, with a spatial PRF of $PRF_{\tx{in}}$ (1/m). For $k\in\ol{0,K-1}$, this yields the DT signal $s'(\cdot)$ (see Fig.~\ref{Fig:UnifiedBlockDiagram}), where
\begin{equation}
  s'(k)
    =s(\nu_r, u_k),\tx{ with }
  u_{k+1}-u_k
    =1/PRF_{\tx{in}}.
  \label{eq:s'k}
\end{equation}

In spotlight mode SAR, unambiguous recovery of the cross-range extent $X_{\tx{in}}$ at boresight (where $\beta=\pi/2$) requires that the spatial domain PRF satisfy $PRF_{\tx{in}}=1/\Delta u_{\tx{in}}\geq PRF_{\tx{in},\min}\equiv (2/\lambda_c)\,\Delta\beta=(2/\lambda_c)\,(X_{\tx{in}}/R)$, where $\Delta u_{\tx{in}}$ and $R$ denote along-track sample spacing (m) and slant range, respectively, $\Delta\beta$ is the beam sweep angle (rad), and $X_{\tx{in}}=R\,\Delta\beta$ is the cross-range extent (m) \cite{Wehner1987, Soumeck1999}. In spotlight mode SAR, $R\,\Delta\beta$ denotes the maximum cross-range extent illuminated by the radar beam; in strip mode SAR, it is denoted by $X_{\tx{in}}=R\,\psi$ where $\psi$ is the real aperture beamwidth.  

\begin{figure*}[!t]
  \centering
  \includegraphics[width=0.85\textwidth]{%
    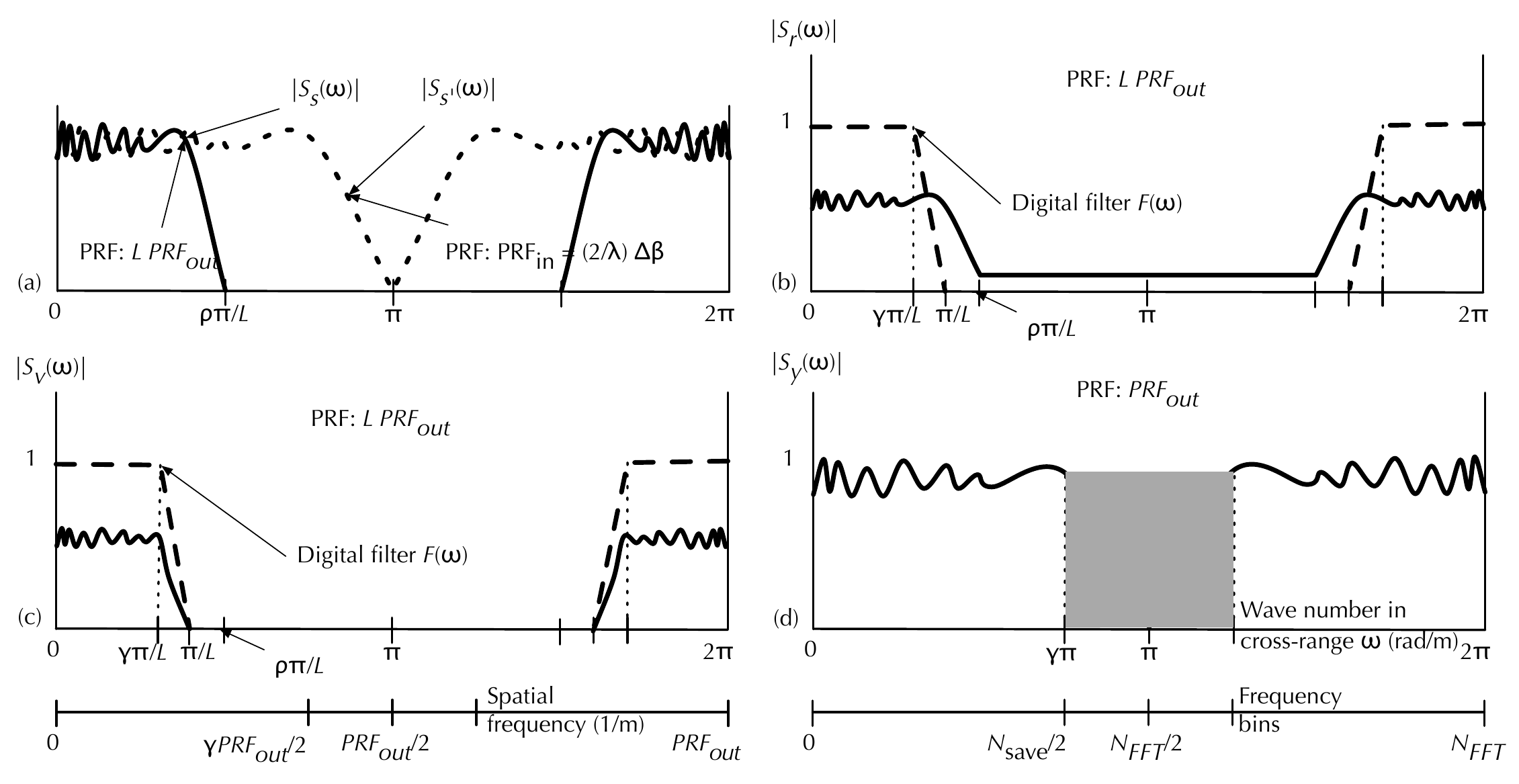}
  \vv\vx
  \caption{\scriptsize{Spatial Doppler bandwidths of: 
  \tb{(a)}~dotted line is $s'(k)$ @ a PRF of $PRF_{\tx{in}}=PRF_{\tx{in},\min}=(2/\lambda_c)\,\Delta\beta$ (1/m) and solid line is $s(n)$ @ a PRF of $L{\cdot}PRF_{\tx{out}}$ (1/m); 
  \tb{(b)}~$r(k)$ @ at a PRF of $L{\cdot}PRF_{\tx{out}}$ (1/m); 
  \tb{(c)}~$v(k)$ @ at a PRF of $L{\cdot}PRF_{\tx{out}}$ (1/m); 
  \tb{(d)}~$y(k)$ @ at a PRF of $PRF_{\tx{out}}$ (1/m).
The digital filter $F(z)$ has its pass and stopband edges at $\gamma\pi/L,\,\gamma<1$, and $\pi/L$, respectively, $\rho=PRF_{\tx{in}}/PRF_{\tx{out}}\geq 1$, $N_{\tx{save}}$ denotes the number of bins guaranteed to be retained in the final image without distortion, and $\gamma=N_{\tx{save}}/N_{FFT}=PBW/PRF_{\tx{out}}<1$. Spatial frequency value (with respect to the PRF $PRF_{\tx{out}}$) and frequency bin axes are also shown underneath \tb{(c)} and \tb{(d)}.}}
  \label{Fig:Y_and_S}
\end{figure*}

Fig.~\ref{Fig:Y_and_S}(a) shows $S_{s'}(\omega_{cr})$, the frequency response of $s'(\cdot)$ when $PRF_{\tx{in}}=PRF_{\tx{in},\min}$. In Fig.~\ref{Fig:Y_and_S} and onwards, we use $\omega$ for $\omega_{cr}$ \cite{Oppenheim1975, Crochiere1981ProcIEEE}.
  \tb{(a)}~A smaller PRF $PRF_{\tx{out}}<PRF_{\tx{in},\min}$ causes aliasing, and the cross-range extent that can be unambiguously recovered is reduced as $X_{\tx{out}}=(\lambda_c/2)\,R\,PRF_{\tx{out}}<R\,\Delta\beta$.
  \tb{(b)}~Uniformly sampling $s(\cdot,\cdot)$ with a higher PRF $L{\cdot} PRF_{\tx{out}},\,L{\in}\mbb{N}_+$, where $L{\cdot}PRF_{\tx{out}}>>PRF_{\tx{in},\min}=(2/\lambda_c)\,\Delta\beta$, yields $s(\cdot)$, where $s(n)=\left.s(\nu_r, u)\right|_{u=n/(L\cdot PRF_{\tx{out}})}$. Its spectrum $S_s(\omega)$ is in Fig.~\ref{Fig:Y_and_S}(b). 

\subsection{Non-Uniformly Sampled Radar Signal}
\label{subsec:nonuniform}


The acquired DT signal $s'(\cdot)$ is a potentially non-uniformly sampled version of $s(\cdot,\cdot)$. The SAR collection receives $s'(\cdot)$ whose variable PRF $PRF_{\tx{in}}$ is assumed to be high enough to sample the available Doppler support for an illuminated cross-range extent of $R\,\Delta\beta$ with no aliasing.  

\bi{Model.} 
We view $s'(k)\leftrightarrow S_{s'}(\omega)$ as being the uniformly densely sampled signal $s(n)\leftrightarrow S_s(\omega)$ but with `missing' samples. Here, $\leftrightarrow$ denotes a DT FT pair (in the deterministic case) or a PSD (power spectral density) pair (in the stochastic case). Then, with appropriate scaling and grid alignment, the signal $r(n)\leftrightarrow S_r(\omega)$ in Fig.~\ref{Fig:UnifiedBlockDiagram} can be viewed as a `gated' version of $s(n)$, i.e., $r(n)=g(n)\,s(n)$, where the gating function $g(n)\leftrightarrow S_g(\omega)$ is a realization of an i.i.d. Bernoulli random process with parameter $p$. So, the probabilities of $g(n)$ taking the values $1$ and $0$ are given by
\begin{equation}
  Pr(g(n)=1)
    =p;\;
  Pr(g(n)=0)
    =1-p,\;\forall\,\mbb{N},
  \label{eq:pBern}
\end{equation}
respectively. Using $S_x(\omega)$ to denote the PSD of the w.s.s. random process $x(\cd)$, we get the PSD of $g(n)$ as
\begin{equation}
  S_g(\omega)
    =p(1-p)+(2\pi p^2)\sum_{k=-\infty}^{+\infty} \delta_D(\omega-2\pi k).
\end{equation}
Since the PSD of $r(n)$ is given by $S_r(\omega)=(1/2\pi)\,(S_g(\omega)\ast S_s(\omega))$, we get $S_r(\omega)$ as a scaled `biased' version of $S_s(\omega)$:
\begin{equation}
  S_r(\omega)
    =\frac{1}{2\pi}\,p(1-p)
     \int_{\omega=-\pi}^{+\pi} S_s(\omega)\,d\omega
       +p^2S_s(\omega).
  \label{eq:bias}
\end{equation}

\bi{Spatial Doppler Bandwidth Recovery at Output PRF.} 
Suppose we are interested in an image signal from which a cross-range extent of $X_{\tx{out}}$ can be recovered with a sampling rate of $PRF_{\tx{out}}$ (1/m). As we show in Appendix~\ref{appA}, such a signal, which approximates a downsampled version of $s(n)$ (which is at the constant PRF $L{\cdot}PRF_{\tx{out}}$), can be generated from the resampling scheme in Fig.~\ref{Fig:UnifiedBlockDiagram} by implementing the following three operations:
  \tb{(a)}~Get $v(n)$ by filtering $r(n)$ by a digital filter $f(n)\leftrightarrow F(z)$ with the magnitude response
\begin{equation}
  |F(\omega)|
    =\begin{cases}
       1,
         & \tx{in the passband $|\omega|\leq\gamma\,\pi/L$}; \\
       0, 
         & \tx{in the stopband $\pi/L\leq|\omega|$},
     \end{cases}
  \label{eq:specs}
\end{equation}
where $L>>1$; 
  \tb{(b)}~normalize $v(n)$ by $f(n)\ast g(n)$ to get $v'(n)$; and finally, 
  \tb{(c)}~$L$-fold decimate $v'(n)$ to get $y(n)$. 

Then, if $p>>\rho/(L+\rho)$, where $\rho=PRF_{\tx{in}}/PRF_{\tx{out}}\geq 1$, the PSD of the output $y(n)$ approximates the PSD of an $L$-fold decimated version of the uniformly densely sampled signal $s(n)$ within the frequency band $[0, \gamma\,\pi]$. So, to recover a spatial Doppler bandwidth of $PBW$ (Hz) corresponding to $N_{\tx{save}}$ bins, we must have $\gamma=N_{\tx{save}}/N_{FFT}$ (or equivalently, $\gamma=PBW/PRF_{\tx{out}}$). Note that $N_{\tx{save}}$ is the number of bins guaranteed to be retained in the final image without distortion. 

More details of the various stages in Fig.~\ref{Fig:UnifiedBlockDiagram} follow next.


\section{Output Grid Spacing Design}
\label{sec:OutputGrid}


Here we select the slow-time output grid spacing $\Delta u_{\tx{out}}$ (m) so that it conforms to a given spatial PRF $PRF_{\tx{out}}$, preserves the resolution in the image domain, and avoids aliasing in both  spatial slow-time and image domains. The output grid spacing $\Delta u_{\tx{out}}/L$ is then used to view the raw data $s'(k)$ (at the variable PRF $PRF_{\tx{in}}$) as being embedded in a uniformly densely sampled signal $r(n)$ (at the constant PRF $L{\cdot}PRF_{\tx{out}}$).


\subsection{Design Steps}


\bi{Fast Fourier Transform (FFT) Size.} 
As commonly practiced in SAR processing, we first select the azimuth compression FFT size $N_{FFT}$ so that read access memory (RAM) and SAR processor power/speed limitations can be met. 

\bi{Synthetic Aperture.} 
Note that $N_{FFT}$ is the next power-of-two FFT size obtained from $N_D$, the number of points necessary to represent the synthetic aperture with no aliasing. Then, the slow-time spectrum oversampling factor is
\begin{equation}
  p_d
    =\dfrac{N_{FFT}\Delta u_{\tx{out}}}{N_D\Delta u_{\tx{out}}}\,K_{cr}
    =\dfrac{N_{FFT}}{N_D}\,K_{cr},
  \label{eq:pdensity}
\end{equation}
where $K_{cr}$ compensates for broadening due to aperture weighting \cite{Jakowatz1996}. Note that $N_D\Delta u_{\tx{out}}$ and $N_{FFT}\Delta u_{\tx{out}}$ denote the slow-time acquisition intervals $D$ and $D_{pad}$ generated by $N_D$ and $N_{FFT}$, respectively. Since $N_D\leq N_{FFT}$, this implies that $K_{cr}\leq p_d$. We used $p_d=1.5$ which generated adequate oversampling to create a visually more pleasing image; oversampling also facilitates the application of certain image processing procedures. We then get $N_D$ from \eqref{eq:pdensity}.

Filtering by the order $N_{pr}$ prototype $H_{pr}(z)$ (see Section~\ref{subsec:design}) requires $N_{pr}$ number of samples to be appended to the $N_D$ number of points laid out along the synthetic aperture. This yields a total of $N_{X\!AC}=N_D+N_{pr}$ number of points.

\bi{Slow-Time Output Grid Spacing and Output.} 
To select the spatial slow-time output grid spacing $\Delta u_{\tx{out}}$, match range and azimuth resolutions to get square radar resolution cells in the oversampled image domain. Equating the range and azimuth resolution expressions in spotlight mode SAR $\delta_r=(c/2)\,(K_r/B_{chirp})$ and $\delta_{cr}=(\lambda_c/2)\,(K_{cr}/\Delta\beta)$ \cite{Wehner1987, Carrara1995, Jakowatz1996}, we get the SAR integration angle $\Delta\beta=(\lambda_cB_{chirp}/c)\,(K_{cr}/K_r)$, which yields the required synthetic aperture length $D$ \cite{Soumeck1999}. The slow-time output grid spacing $\Delta u_{\tx{out}}$ and the corresponding spatial sampling frequency $PRF_{\tx{out}}$ of the resampled data are then given by $\Delta u_{\tx{out}}=D/N_D$ so that $PRF_{\tx{out}}=1/\Delta u_{\tx{out}}$.

\bi{Cross-Range Extent.} 
With $X_{\tx{out}}=(\lambda_c/2)\,R\,PRF_{\tx{out}}$, and for the selected FFT size $N_{FFT}$, we are attempting to fit as much cross-range extent $X_{\tx{out}}$ as possible so that PRF conversion is more efficient: it is more parallel processing friendly, and aid the SAR processor to partition the image into smaller patches of cross-range extent allowing smaller FFT sizes to be run faster on parallel nodes. 


\subsection{Grid Alignment}
\label{subsec:3c}

\bi{Scaling.} 
We first utilize a linear transformation to map the input pulses spatial information into the output spatial grid. To explain, let $s(\nu_r, \alpha)=s(\nu_r, u)|_{u=\Phi^{-1}(\alpha)}$, where
\begin{equation}
  \alpha
    =\Phi(u) 
    =(u-u_{\tx{mid}})/\Delta u_{\tx{out}}+(N_D+1)/2.
  \label{eq:alpha}
\end{equation} 
Here, $u\in [u_{\min},u_{\max}]$, $\Delta u_{\tx{out}}=(u_{\max}{-}u_{\min})/N_{X\!AC}$ is the output spacing along-track, and $N_D=N_{X\!AC}-N_{pr}$ is the number of output points along the acquisition interval $D$. With $\alpha_k=\Phi(u_k),\,k\in\ol{0,K}$, this transformation in \eqref{eq:alpha} transforms the sequence $s'(\cdot)$ to the sequence $s''(\cdot)$, where 
\begin{equation}
  s''(k)
    \equiv s(\nu_r, \alpha_k)
    =s(\nu_r, u_k)
    \equiv
     s'(k),\;
  \alpha_k
    =\Phi(u_k).
  \label{eq:shat}
\end{equation}

\bi{Grid Alignment.} 
The next step involves aligning the sampled values in $s''(\cdot)$ onto a dense grid corresponding to the rate $L{\cdot}PRF_{\tx{out}}$. This creates the sequence $r(\cdot)$, where 
\begin{equation}
  r(n)
    =s''(\alpha_k),
     \tx{ for }
     n=\lfloor L\cdot\alpha_k\rfloor,
  \label{eq:input_rescale}
\end{equation}
and $r(n)=0$ otherwise. Here, $\lfloor{x/L}\rfloor=(x-(x)_L)/L$, where $(x)_L$ denotes the remainder when $x$ and $L$ are the dividend and divisor, respectively. This aligns each sampled value in $s''(k)$ to a grid point (within the densely sampled grid with rate $L{\cdot}PRF_{\tx{out}}$) which is closest but not higher than $\alpha_k$. The remaining grid points (i.e., the `missing' samples) have value $0$. So, $r(\cdot)$ can be viewed as a `gated' version of $s(\cdot)$, where only some samples of $s(\cdot)$ appear in $r(\cdot)$ while the others take values $0$ (see Section~\ref{subsec:nonuniform} and in Appendix~\ref{appA}).


\section{Filter Design and Implementation}
\label{sec:filter}


POLYPHASE requires a narrowband digital filter (see Claim~\ref{cl:App} in Appendix~\ref{appA} and  \eqref{eq:specs}). A flexible SAR focusing processor (i.e., TerraSAR-X) must accommodate a wide range of integer and non-integer resample ratios (which can  vary with radar collection parameters, geometry, processed image resolution and scene size, FFT size, etc.). This calls for a single filter which is sufficiently flexible to handle a large range of resampling ratios and can be efficiently implemented. A polyphase implementation of the filter addresses both these issues. Polyphase architectures, a critical component in multirate digital systems, are computationally more efficient because they operate at the lowest sampling rate \cite{Crochiere1981ProcIEEE}. 


\subsection{Filter Design}
\label{subsec:design}


\bi{Step~1. Prototype Filter.} 
Use a standard finite impulse response (FIR) filter design technique (e.g., the Remez exchange/Parks-McClellan algorithm \cite{Oppenheim1975}) to design a `prototype' $F_{pr}(z)=\sum_{m=0}^{N_{pr}} f_{pr}(m)\,z^{-m}$ with pass and stopband edge frequencies located $L$-times the values specified in \eqref{eq:specs}:
\begin{equation}
  F_{pr}(\omega)
    =\begin{cases}
       1,
         & \tx{for $\omega\in [0, \gamma\,\pi]$;} \\
       0,
         & \tx{for $\omega = \pi$}.
     \end{cases}
  \label{eq:Prototype Filter}
\end{equation}
We choose a Type-II FIR filter design (with symmetric filter taps and odd filter order $N_{pr}$). One may also use a Type-I FIR filter (with symmetric filter taps and even filter order).

\bi{Step~2. Shaping Filter.} 
As is typical in interpolated FIR (IFIR) filter design \cite{Saramaki1988ToCS, Lyons2003SPM}, generate an $L$-fold upsampled version of $f_{pr}(n)$ to get the `shaping' filter \begin{equation}
  f_{be}(n)
    =\begin{cases}
       f_{pr}(n/L),
         & \tx{for $n=0, L, \ldots, N_{pr}L$}; \\
       0,
         & \tx{otherwise}.
     \end{cases}
  \label{eq:Band Edge Shaping Filter}
\end{equation}
so that $F_{be}(z)=F_{pr}(z^L)$. This $N_{pr}L$-order filter's frequency response is the desired response in \eqref{eq:specs}, except that spectral `images' of this desired response now appear within the Nyquist interval. The filter order $N_{pr}L$ enables a polyphase design consisting of $L$ sub-filters \cite{Oppenheim1975}.

\bi{Step~3. Image Suppression.} 
IFIR designs requires a `masking' filter to suppress these extra images \cite{Lyons2003SPM}. But, this increases the length of our overall impulse response (IPR) and the filter order beyond $N_{pr}L$. So, we employ a direct least squared integral error (LSIE) FIR design \cite{Parks1987, Laakso1996SPM} to design an $N_{pr}L$-order filter to approximate $F_{be}(\omega)$ in the frequency interval $[0, \pi/L]$. Then, the `ideal' frequency response to be approximated is $F_{id}(\omega)=F_{be}(\omega)\,F_{LPF}(\omega)$, where 
\begin{equation}
  F_{LPF}(\omega)
    =\begin{cases}
       1,
         & \tx{for $|\omega|\leq\pi/L$}; \\
       0,
         & \tx{otherwise}.
     \end{cases}
\end{equation}
Note that $F_{LPF}(\omega)\leftrightarrow f_{LPF}(n)=(1/L)\,\tx{sinc}(\pi n/L)$ and $f_{id}(n)=f_{be}(n)\ast f_{LPF}(n)$. So, 
\begin{equation}
  f_{id}(n)
    =\frac{1}{L}
     \sum_{m=0}^{N_{pr}} 
     f_{pr}(m)\,
     \tx{sinc}
     \left(
       (n-mL)\frac{\pi}{L}
     \right),
\end{equation}
and the filter with support in $[0, N_{pr}L]$ which minimizes the LSIE with $F_{id}(\omega)$ is \cite{Parks1987, Laakso1996SPM}
\begin{equation}
  f(n)
    =\frac{1}{L}
     \sum_{m=0}^{N_{pr}} f_{pr}(m)\,
     \tx{sinc}
     \left(
       (n-mL)\frac{\pi}{L}
     \right).
\end{equation}

\begin{figure}[!t]
  \begin{center}
  \subfigure[Polyphase representation of $F(z)$.]{%
    \includegraphics[width=0.75\linewidth]{%
      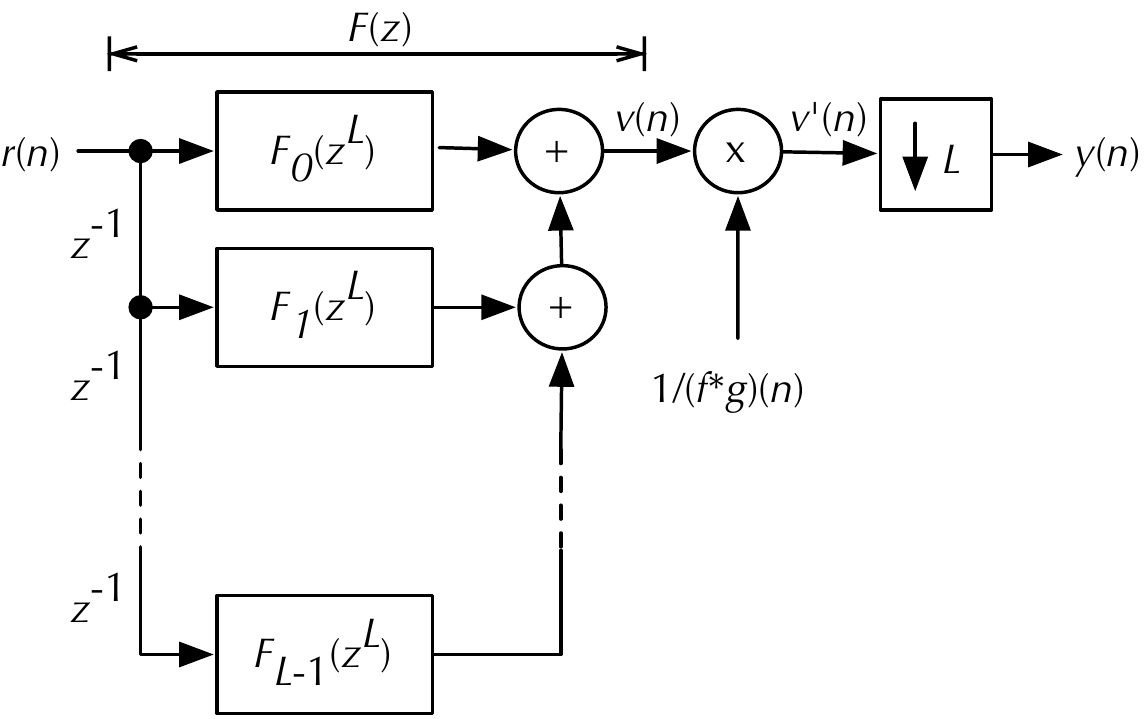}} \\
  \subfigure[Efficient implementation of (a).]{%
    \includegraphics[width=0.75\linewidth]{%
      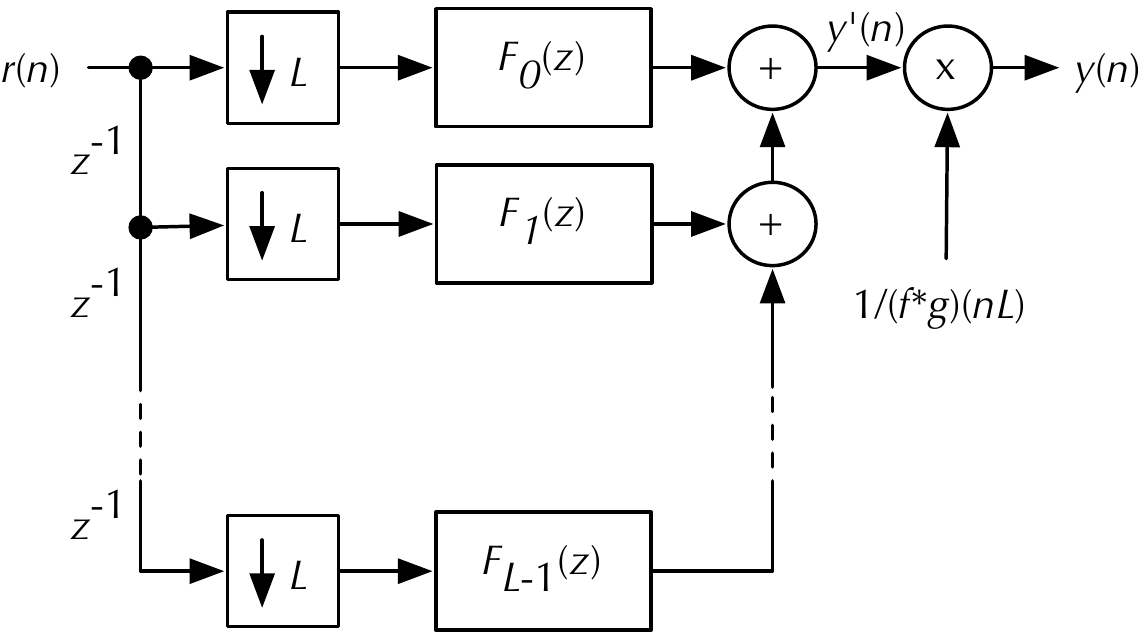}}
  \end{center}
  \vv\vx
  \caption{\scriptsize{Efficient polyphase implementation of the filter $F(z)$. The structure in (b) allows the digital filtering operations in (a) to be carried out at the lowest sampling rate.}}
  \label{fig:Polyphase}
\end{figure}


\subsection{Polyphase Filter Implementation}


Consider the $L$-fold polyphase representation of $F(z)$ \cite{Oppenheim1975}:
\begin{equation}
  F(z)
    =\sum_{\ell=0}^{L-1} z^{-\ell}F_{\ell}(z^L),\tx{ with }
  F_{\ell}(z)
    =\sum_{n=0}^{N_{pr}} f_{\ell}(n)\,z^{-n}.
  \label{eq:F_l}
\end{equation}
Here, $f_{\ell}(n)=f(nL+\ell)$. See Fig.~\ref{fig:Polyphase}(a). Note that, $F_0(z)$ is of order $N_{pr}$; $F_{\ell}(z),\,\ell\in\ol{1,L-1}$, is of order $N_{pr}-1$. 

To create the final output $y(n)$, we normalize the filtered output $v(n)$ to get $v'(n)=v(n)/(f(n)\ast g(n))\vert_{nL}$, and $L$-fold downsample $v'(n)$. See Fig.~\ref{Fig:UnifiedBlockDiagram} and Fig.~\ref{fig:Polyphase}(a). At this juncture, we make use of a well-known alternate implementation of the structure in Fig.~\ref{fig:Polyphase}(a), viz., Fig.~\ref{fig:Polyphase}(b), which allows the filtering operations to be performed at the lowest sampling rate \cite{Crochiere1981ProcIEEE}. The input to each polyphase component $F_{\ell}(z)$ is now a downsampled (and shifted) version of the input $r(\cdot)$. 

\bi{Normalization.} 
In Fig.~\ref{fig:Polyphase}(a), the normalized output $v'(n)$ must be downsampled to produce the required output $y(n)$. In the more efficient implementation of Fig.~\ref{fig:Polyphase}(b), normalization immediately produces $y(n)$. This `normalized' convolution may be interpreted as associating a level of `confidence' with the received signal \cite{knutsson1993normalized, westin1994equivalence, andersson2002continuous}. For example, each missing sample is associated with zero confidence. Accordingly, $f(n)\ast g(n)$, where the gating function $g(n)$ (see Section~\ref{subsec:nonuniform}) captures the confidence associated with the received signal $r(\cdot)$. The normalization factor $(f(n)\ast g(n))\vert_{nL}$ is nothing more than the sum of the filter coefficients that contribute to the output computation. In actual implementation, $(f(n)\ast g(n))\vert_{nL}$ is computed by maintaining a separate buffer which accumulates the sum of the coefficients used at each output location. 

\bi{Generating the Output.} 
With $n_{pr}=n-N_{pr}$, we get the output $y(n)=y'(n)/(f(n)\ast g(n))\vert_{nL}$ as  
\begin{align}
  y'(n)
    &=\sum_{m=n_{pr}}^n f_0(n-m)\,r(mL)
      \notag \\
    &\quad
        +\sum_{\ell=1}^{L-1}
         \sum_{m=n_{pr}+1}^n 
         f_{\ell}(n-m)\,r(mL-\ell)
      \notag \\
    &=f_0(N_{pr})\,r(n_{pr}L)
      \notag \\
    &\quad
        +\!
         \sum_{m=n_{pr}+1}^n 
         \sum_{\ell=mL-(L-1)}^{mL} 
         f_{mL-\ell}(n-m)\,r(\ell).
  \label{eq:exp2}
\end{align}
So, computation of one output sample $y'(n)$ requires all non-zero input samples $r(\ell)$ s.t. $(n-N_{pr})L\leq\ell\leq nL$. Conversely, the single non-zero input sample $r(\ell)$ affects the computation of all output samples $y(n)$ s.t. $\ell\leq nL\leq \ell+N_{pr}L$, i.e., 
\begin{equation}
  y(n) 
  \tx{ s.t. } 
  \left\lfloor{%
    \frac{\ell-1}{L}}
  \right\rfloor+1
    \leq n
    \leq
     \left\lfloor{%
       \frac{\ell}{L}}
     \right\rfloor+N_{pr}.
  \label{eq:ell2n}
\end{equation}

Note that, with the polyphase implementation, a given non-zero input sample $r(\ell^*)$ s.t. $(n-N_{pr})L\leq\ell^*\leq nL$, is operated on by only one polyphase component $f_x(\cd)$, where $x=m^*L-\ell^*$ with $m^*=\left\lfloor{(\ell^*-1)/L}\right\rfloor+1$ (see Claim~\ref{cl:xm} in Appendix~\ref{appB}).     

\bi{Summary.} 
We use Fig.~\ref{fig:PolyphaseImplementation} to explain the above operations.

  \tb{(a):}~\emph{Black} circles denote $s'(\cdot)$, the non-uniformly sampled input SAR signal (see \eqref{eq:s'k}). 

  \tb{(b):}~\emph{Grey} circles denote $s(\cdot)$, the signal that is densely uniformly sampled at a rate of $L\,PRF_{\tx{in}}$ (see Section~\ref{subsec:sn}). Note that $s(\cdot)$ is unavailable, and the input $s'(\cdot)$ is viewed as being generated from $s(\cdot)$, but with a high fraction of $s(\cdot)$'s samples missing; the quantity $p$ is an approximation of the fraction of samples that is not missing (see \eqref{eq:pBern}). 
  
  \tb{(c):}~\emph{Black} circles denote $r(\cdot)$, which is the input $s'(\cdot)$ aligned to the same `sampling grid' as $s(\cdot)$ (see \eqref{eq:input_rescale}). To explain, let $N_{pr}=5$ and $L=3$ (our actual implementation uses $L=64$). So, the order of the narrowband digital filter $F(z)$ is $N_{pr}L=15$ and it has $L=3$ polyphase components $\{F_0(z), F_1(z), F_2(z)\}$: $F_0(z)$ is of order $N_{pr}=5$; $F_1(z)$ and $F_2(z)$ are each of order $N_{pr}-1=4$ (see \eqref{eq:specs}, \eqref{eq:Prototype Filter}, and \eqref{eq:F_l}). 
  
  \emph{Ringed black} circle identifies one non-zero input sample $r(\ell^*),\,\ell^*=8$. With $L=3$, the range of $\ell$ in the second summation in \eqref{eq:exp2} is $3m\leq\ell\leq 3m-2$. In this range, $\ell=\ell^*=8$ occurs only when $8=3m^*-1$, i.e., when $m^*=3$. In turn, from the first summation in \eqref{eq:exp2}, the only values of $n$ that would require $r(\ell^*)=r(8)$ must satisfy $n-4\leq m^*=3\leq n$, i.e., $3\leq n\leq 7$ (as \eqref{eq:ell2n} indicates). So, when the input sample $r(\ell^*)=r(8)$ is received, we must update the output samples $y'(n),\, 3\leq n\leq 7$, by $f_{3m^*-\ell^*}(n-m^*)\,r(\ell^*)=f_1(n-3)\,r(8)$. 
  
\begin{figure*}[htpb]
  \begin{center}
  \includegraphics[width=0.65\linewidth]{%
    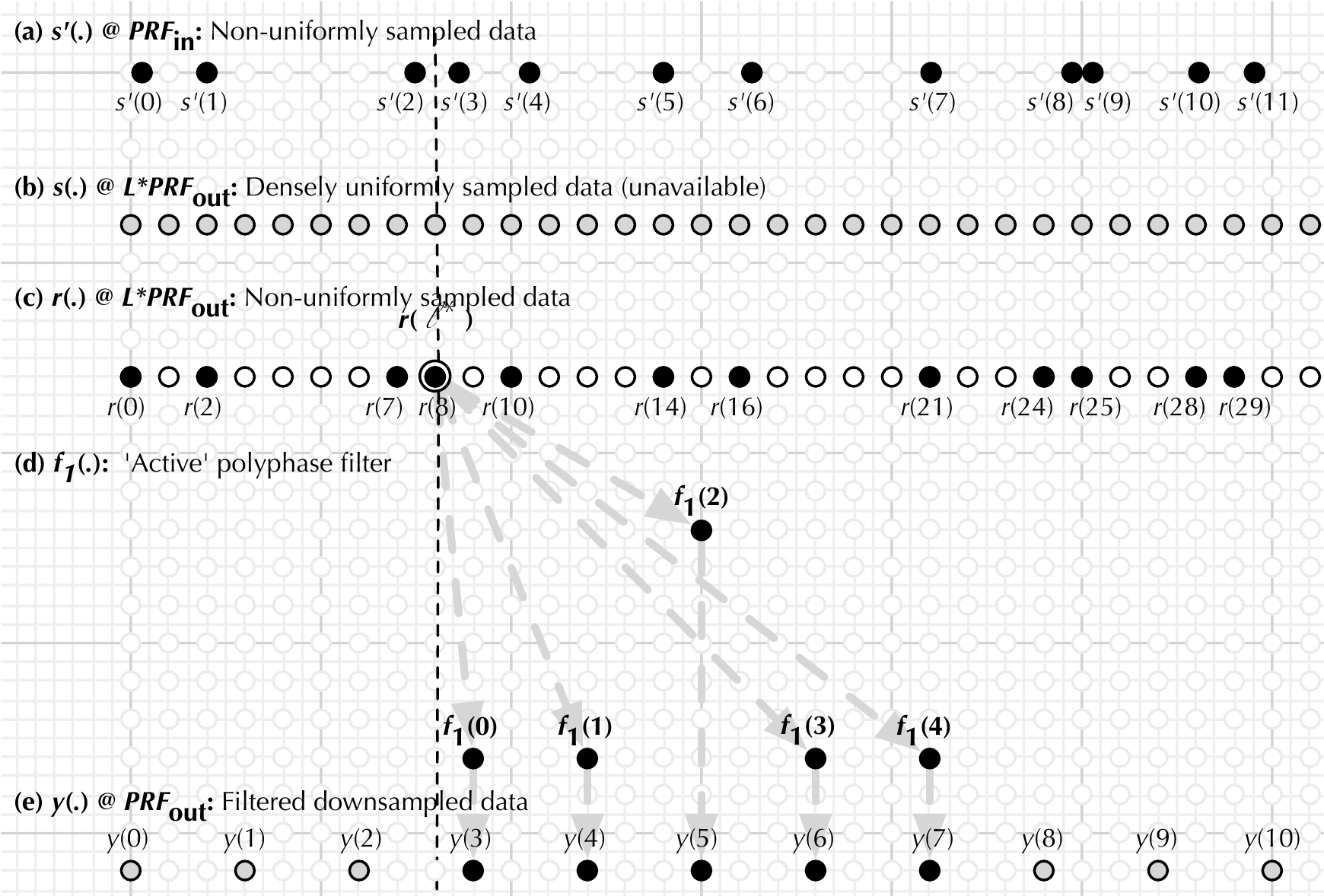}
  \end{center}
  \vv\vx
  \caption{\scriptsize{Implementation of POLYPHASE with $N_{pr}=5$ and $L=3$. See Figs~\ref{Fig:UnifiedBlockDiagram} and \ref{fig:Polyphase}. The only polyphase component that operates on the input sample $r(\ell^*)=r(8)$ is $f_1(n)\leftrightarrow F_1(z)$ which is of order $N_{pr}-1=4$ and has $N_{pr}=5$ taps. The filtered outputs $f_1(n-3)\,r(\ell^*),\,n\in\ol{3,7}$, are used to update the output sample $y(n)$.}}
  \label{fig:PolyphaseImplementation}
\end{figure*}

  \tb{(d):}~$\ell^*=8$ implies $m^*=3$. So, \emph{only} $f_{3m^*-\ell^*}(\cdot)=f_1(\cdot)$ polyphase component is activated (Claim~\ref{cl:xm} in Appendix~\ref{appB}). \emph{Black} circles indicate the $N_{pr}=5$ taps of the filter $f_1(n)\leftrightarrow F_1(z)$. The \emph{gray} arrows show output samples being updated by these taps operating on $r(\ell^*)$. 
  
  \tb{(e):}~\emph{Black} circles denote $y(\cdot),\,3\leq n\leq 7$, the output samples that get updated with the input sample $r(\ell^*)=r(8)$. This updating is carried out `on-the-fly' with no input buffering. 
  
\emph{Gray} circles denote the full set of output samples that need to be computed. Its PSD approximates the PSD of an $L$-fold  decimated ($L=3$) version of $s(n)$ within the frequency band $[0, \gamma\,\pi]$ (Claim~\ref{cl:App} in Appendix~\ref{appA}).

Fig.~\ref{Fig:Input_Center} shows where the proposed scheme belongs within the image formation process. 
\begin{figure}[!h]
  \centering
  \includegraphics[width=0.85\linewidth]{%
    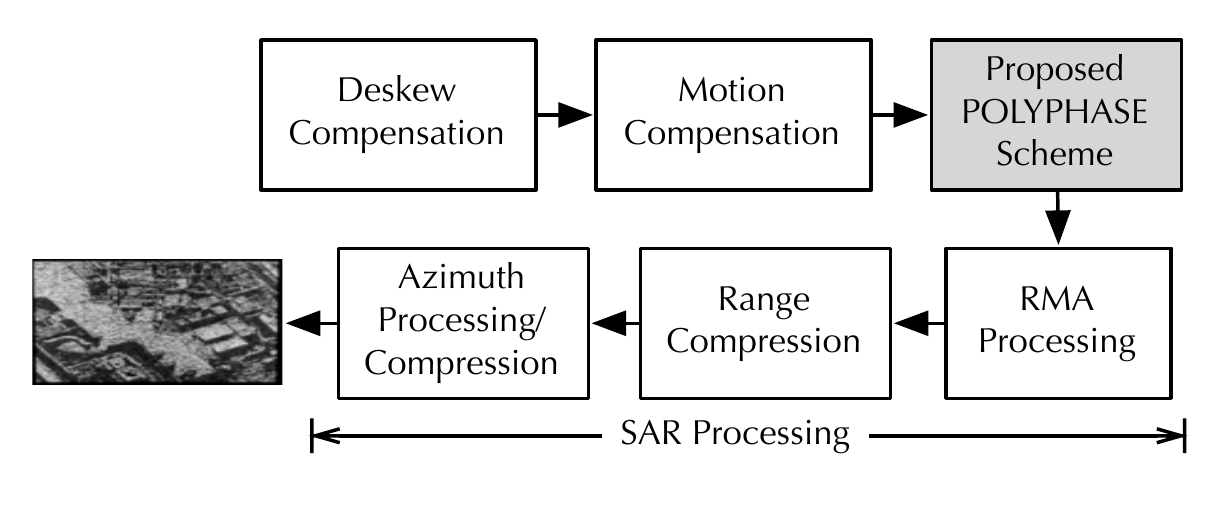}
  \vv\vx
  \caption{\scriptsize{The proposed POLYPHASE scheme's role within the image formation process.}}
  \label{Fig:Input_Center}
\end{figure}


\section{Results}
\label{sec:Results}


\subsection{Synthetic Data}
\label{sec:SimulatedData}


\begin{figure*}[htpb]
  \begin{center}
  \subfigure[Slow PRI variation.]{%
    \includegraphics[width=0.3\linewidth]{%
      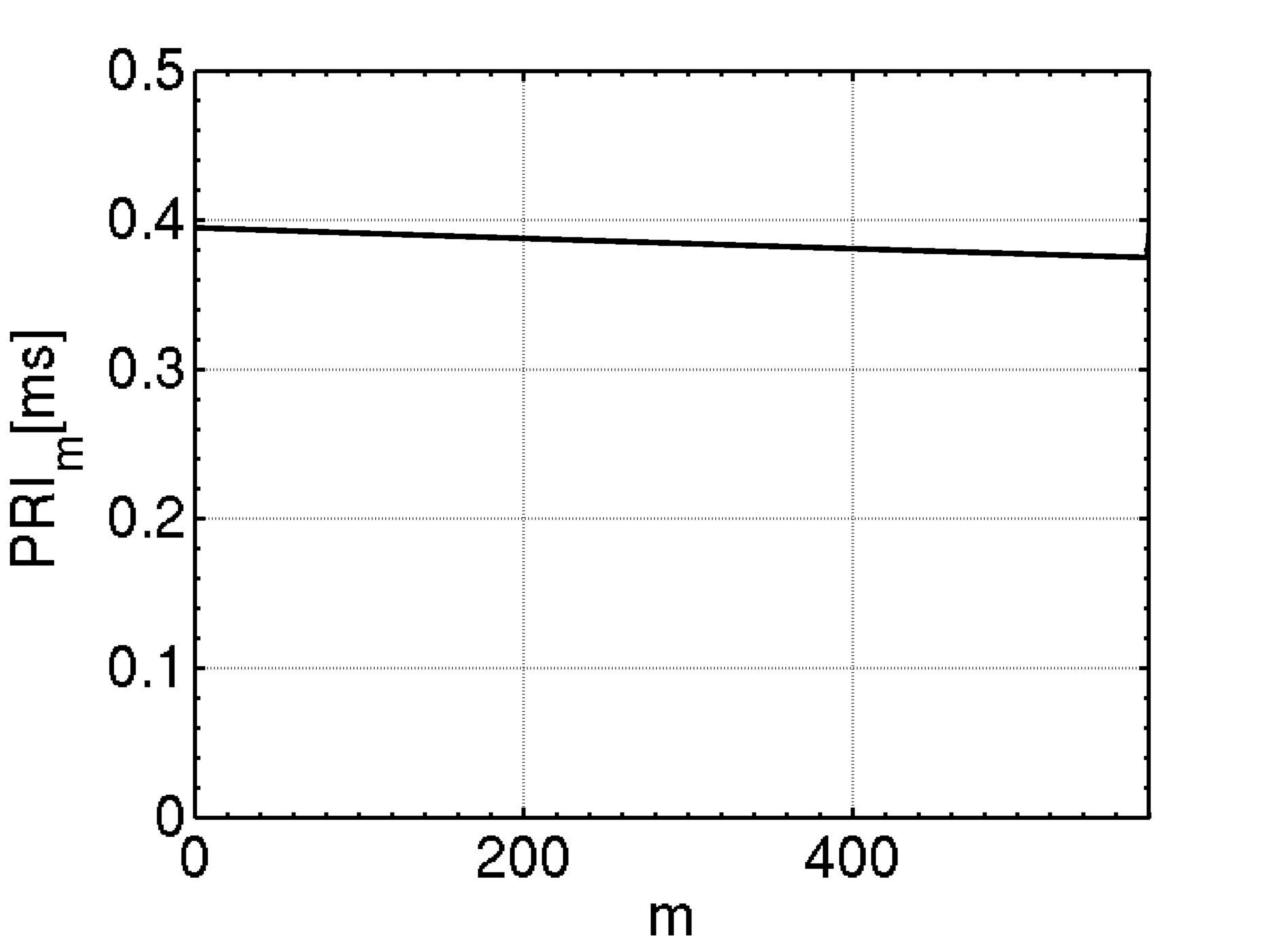}}
  \subfigure[Fast PRI variation.]{%
    \includegraphics[width=0.3\linewidth]{%
      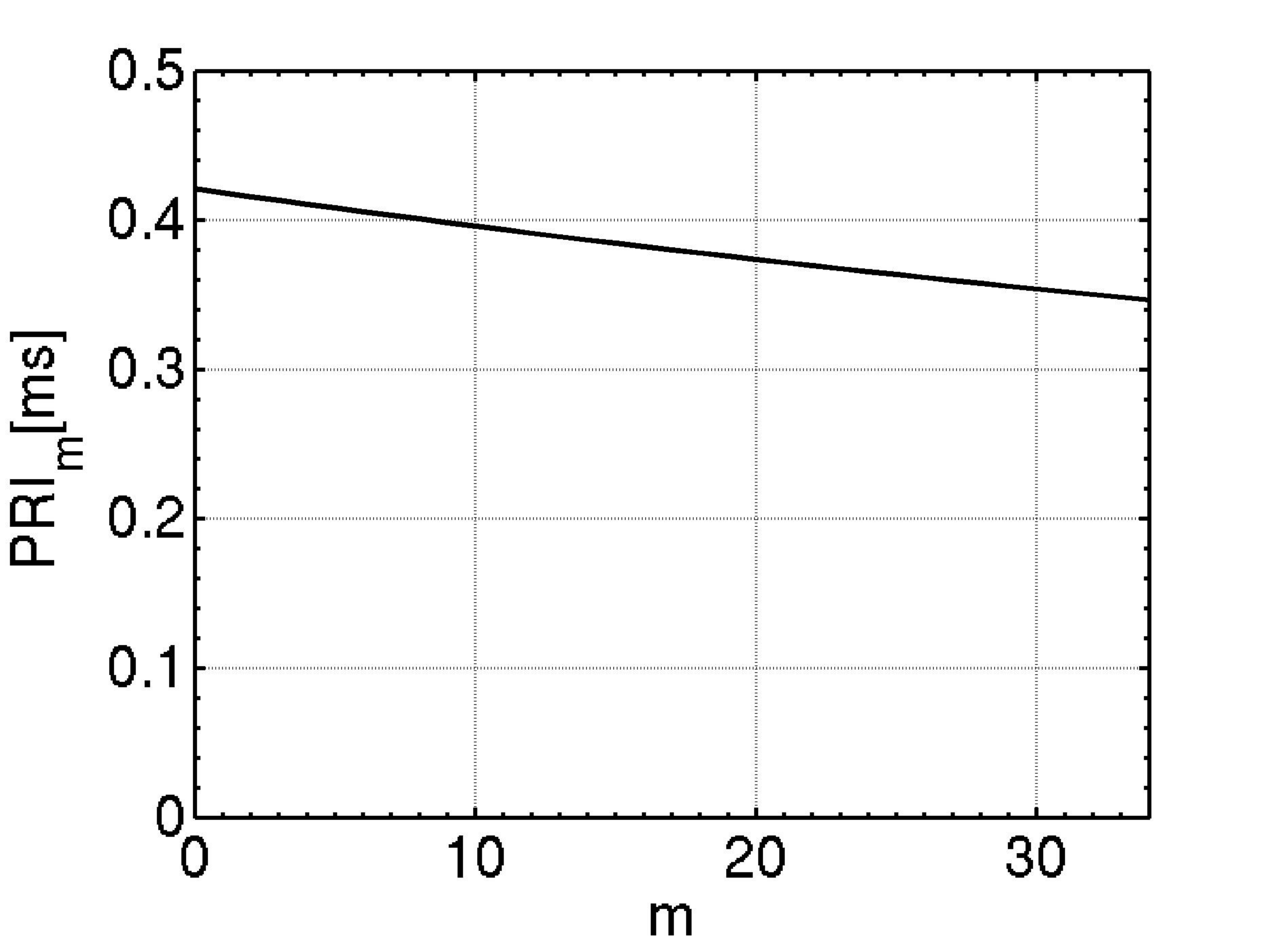}}
  \subfigure[Elaborate PRI variation.]{%
    \includegraphics[width=0.3\linewidth]{%
      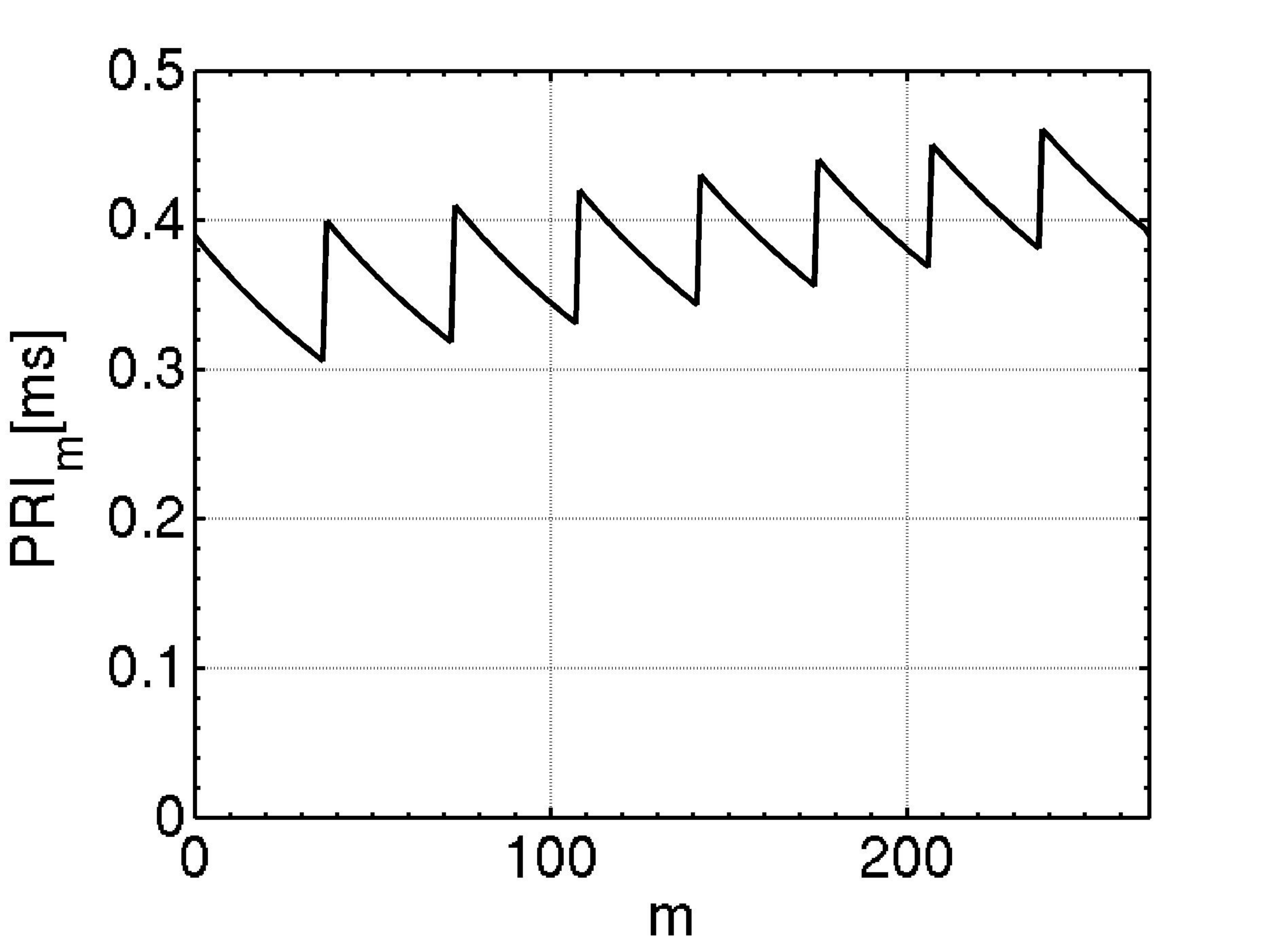}}
  \end{center}
  \vv\vx
  \caption{\scriptsize{One period of the PRI variations. Each variation has mean PRI $PRI_{\tx{mean}}=0.385$ (ms), minimum range of swath $R_{0_{\min}}=837$ (km), maximum range of swath $R_{0_{\max}}=1047$ (km), and pulse width $15$ ($\mu$s) \cite{Villano2014ToGRS}. 
    (a)~Slow PRI variation with $PRI_{\max}=0.395$ (ms), $PRI_{\min}=0.375$ (ms), period $580$ (m). 
    (b)~Fast PRI variation with $PRI_{\max}=0.421$ (ms), $PRI_{\min}=0.349$ (ms), period $34$ (m). 
    (c)~Elaborate PRI variation with $PRI_{\max}=0.461$ (ms), $PRI_{\min}=0.309$ (ms), period $268$ (m).}}
  \label{fig:PRISequences}
\end{figure*}

Here we employ the PRI variations in \cite{Villano2014ToGRS}, viz., the slow, the fast, and the elaborate PRI variations in Fig.~\ref{fig:PRISequences}. To see how the azimuth IPR is affected by these different PRI variations, two different scenarios, each containing 3 scatterers and covering a total unambiguous azimuth extent of $40$ (km), are considered. In Scenario~I, the 3 scatterers are located at $\{-17, 0, +17\}$ (km); in Scenario~II, the 3 scatterers are located at $\{-175, 0, +175\}$ (m). We used a slant range of $R=1000$ (km), an orbit height of $760$ (km), a wavelength of $\lambda=0.2384$ (m) (L-band), and a planar antenna of length of $7$ (m). The azimuth processed bandwidth is set to $PBW=800$ (Hz). In addition, an azimuth Hamming window ($\alpha = 0.6$) and a window which compensates for the azimuth antenna pattern are employed to arrive at an azimuth resolution of $7$ (m). The output PRI for all simulations is $PRI_{\tx{out}}=0.417$ (ms). 

\begin{figure*}[htpb]
  \begin{center}
  \hspace*{-0.2in}
  \subfigure[BLUI with slow PRI variation.]{%
    \includegraphics[width=0.53\linewidth]{%
      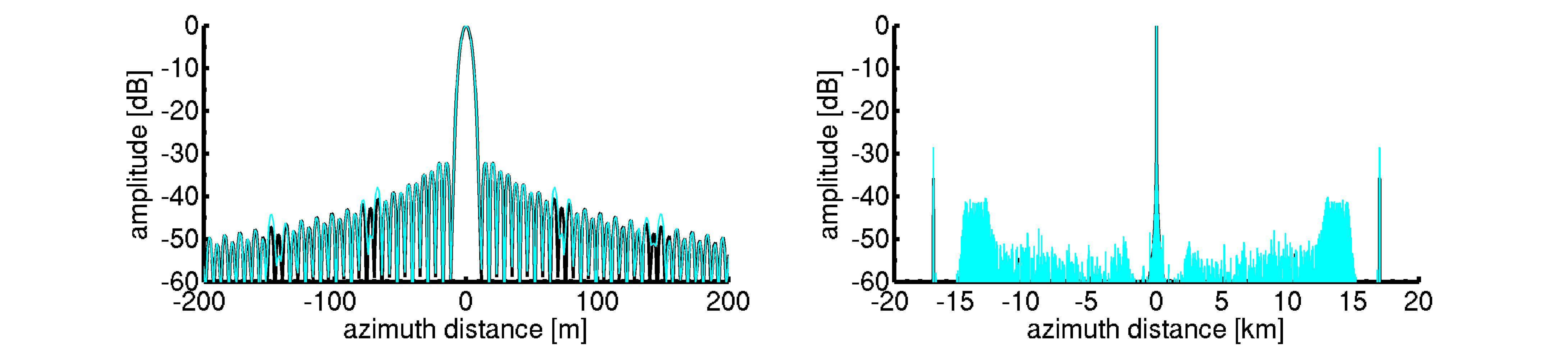}}\hspace*{-0.4in}
  \subfigure[POLYPHASE with slow PRI variation.]{%
    \includegraphics[width=0.53\linewidth]{%
      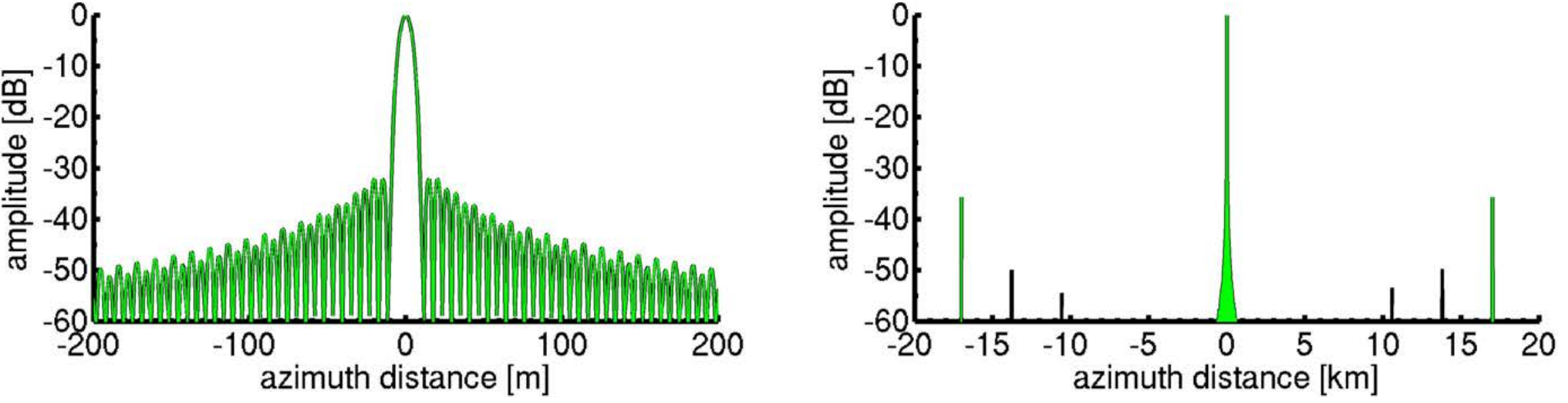}}
  \hspace*{-0.2in}
  \subfigure[BLUI with fast PRI variation.]{%
    \includegraphics[width=0.53\linewidth]{%
      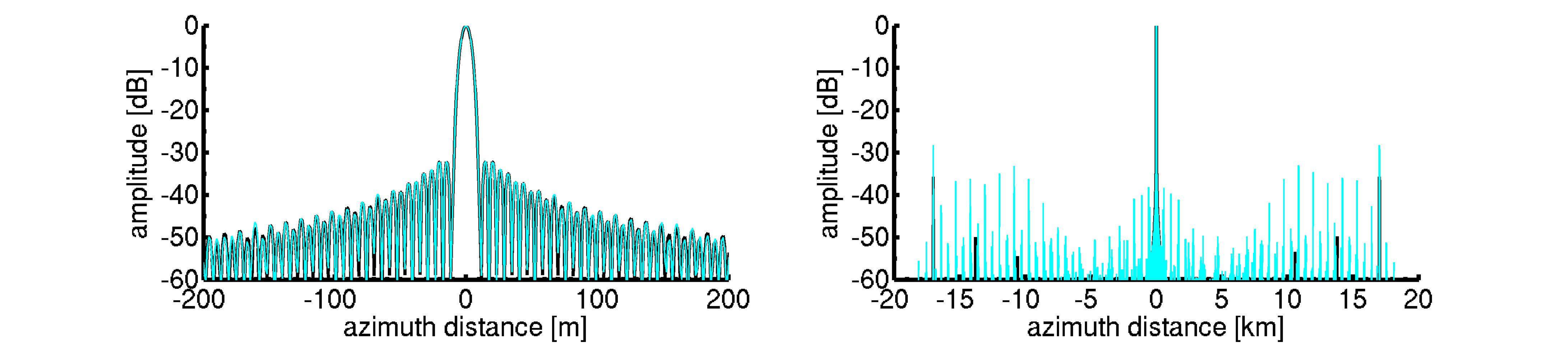}}\hspace*{-0.4in}
  \subfigure[POLYPHASE with fast PRI variation.]{%
    \includegraphics[width=0.53\linewidth]{%
      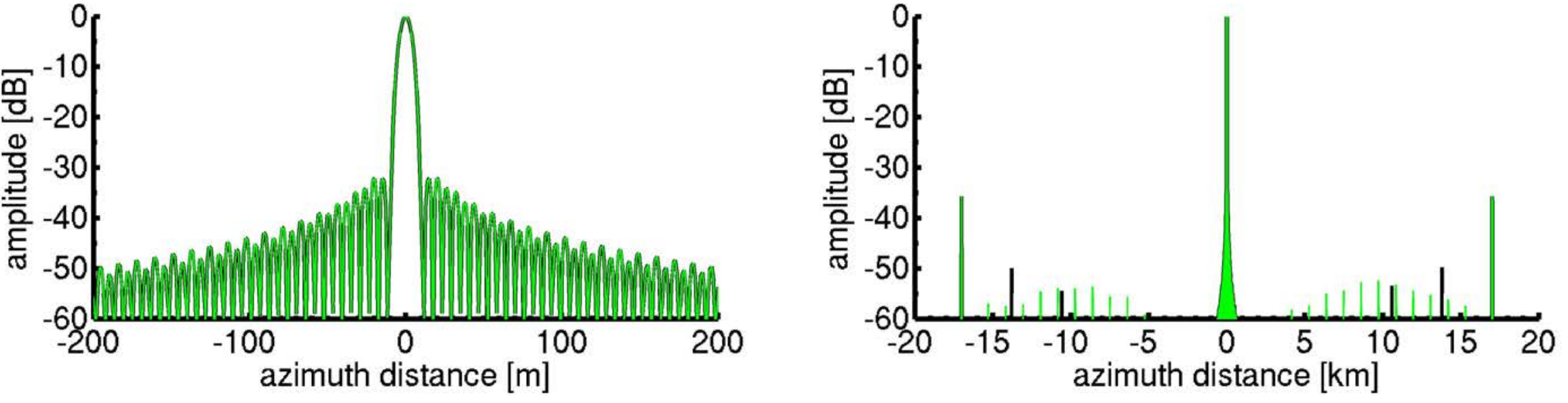}}      
  \hspace*{-0.2in}
  \subfigure[BLUI with elaborate PRI variation.]{%
    \includegraphics[width=0.53\linewidth]{%
      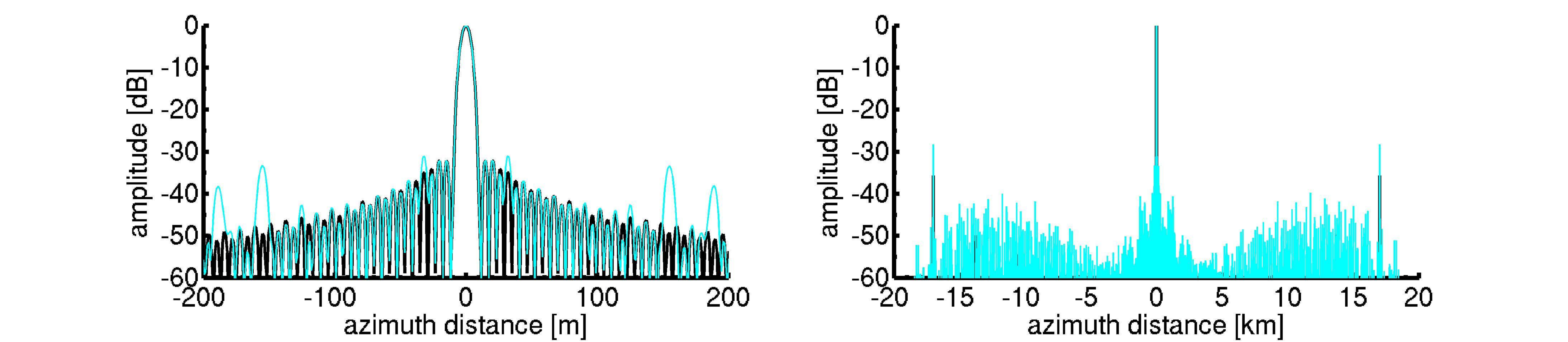}}\hspace*{-0.4in}
  \subfigure[POLYPHASE with elaborate PRI variation.]{%
    \includegraphics[width=0.53\linewidth]{%
      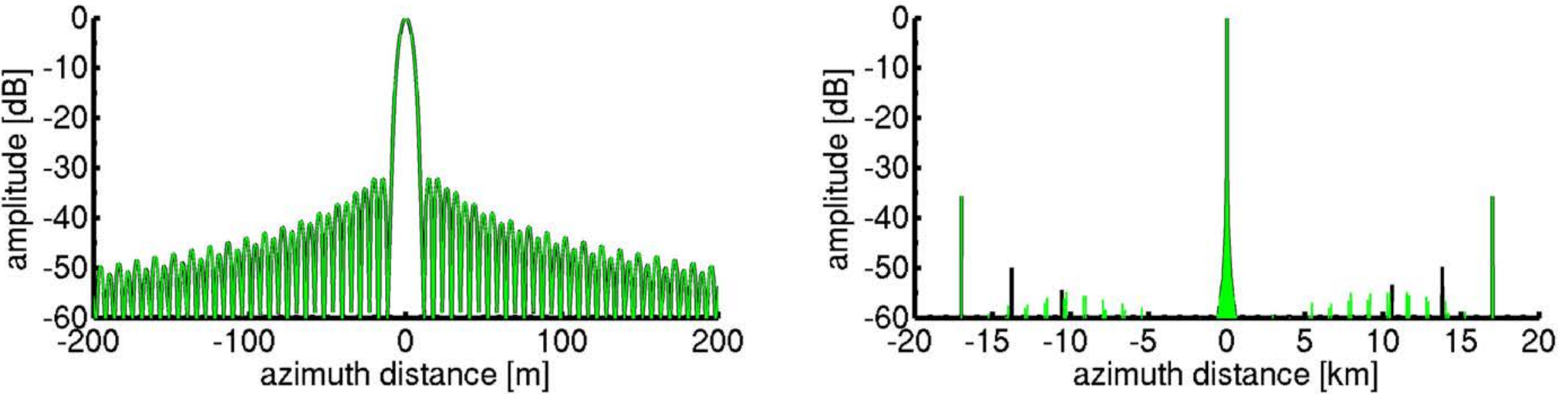}}      
  \end{center}
  \vv\vx
  \caption{\scriptsize{Scenario~I (scatterers at $\{-17, 0, +17\}$ (km)): Comparison of BLUI and POLYPHASE using the azimuth IPRs corresponding to the 3 different PRI variations in Fig.~\ref{fig:PRISequences}. Each IPR is depicted  over $400$ (m) and $40$ (km) azimuth scales. The underlying black curves represent the reference IPR of a SAR system with a constant PRI  equal to the mean PRI $PRI_{\tx{mean}}=0.385$ (ms).}}
  \label{Fig:Performance}
\end{figure*}

\bi{Scenario~I (Scatterers at $\{-17, 0, +17\}$ (km)).}  
Fig.~\ref{Fig:Performance} shows the azimuth IPRs generated by BLUI and POLYPHASE for the 3 periodic PRI variations in Fig.~\ref{fig:PRISequences} with scatterers at $\{-17, 0, +17\}$ (km). The underlying black curves represent the reference IPR of a SAR system with a constant PRI equal to the mean PRI $PRI_{\tx{mean}}=0.385$ (ms).

\begin{figure*}[htpb]
  \begin{center}
  \hspace*{-0.2in}
  \subfigure[BLUI with slow PRI variation.]{%
    \includegraphics[width=0.53\linewidth]{%
      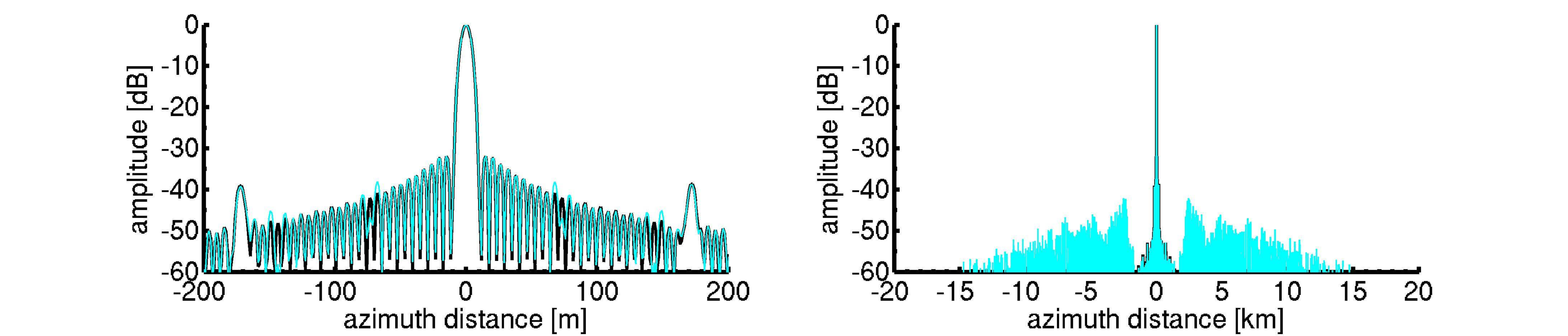}}\hspace*{-0.4in}
  \subfigure[POLYPHASE with slow PRI variation.]{%
    \includegraphics[width=0.53\linewidth]{%
      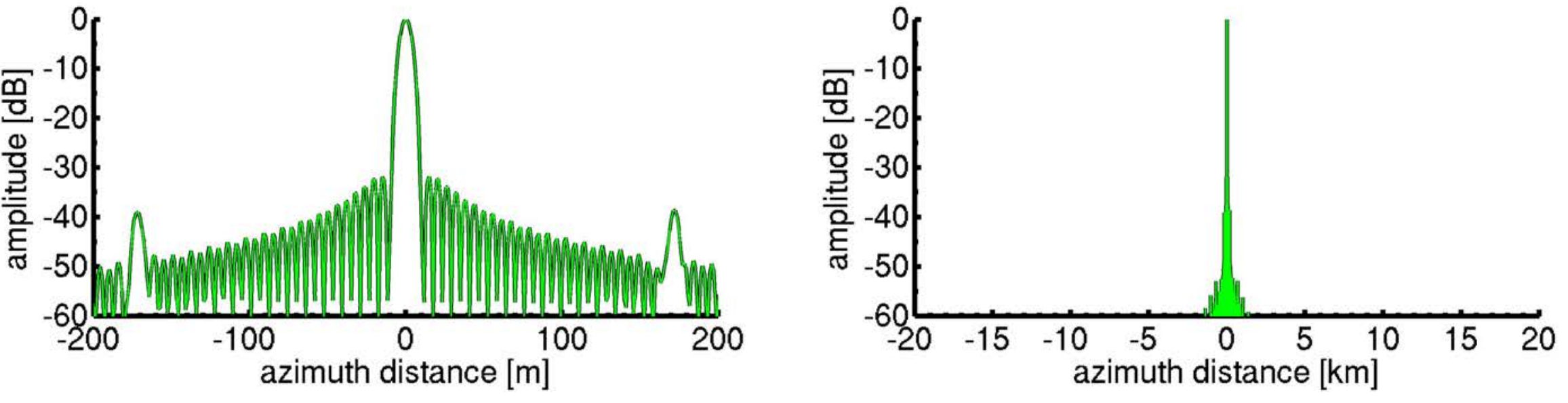}}
  \hspace*{-0.2in}
  \subfigure[BLUI with fast PRI variation.]{%
    \includegraphics[width=0.53\linewidth]{%
      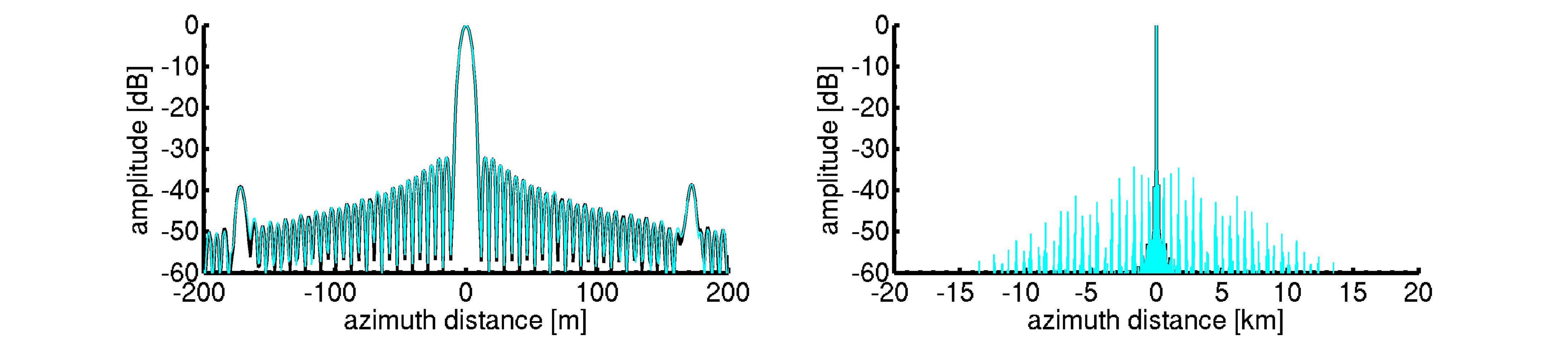}}\hspace*{-0.4in}
  \subfigure[POLYPHASE with fast PRI variation.]{%
    \includegraphics[width=0.53\linewidth]{%
      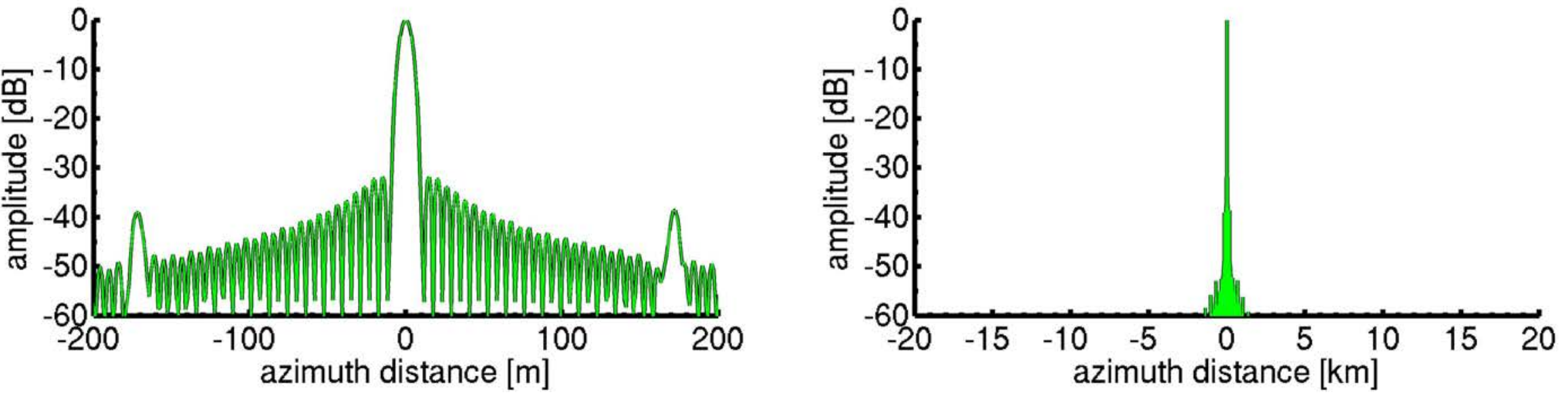}}
  \hspace*{-0.2in}
  \subfigure[BLUI with elaborate PRI variation.]{%
    \includegraphics[width=0.53\linewidth]{%
      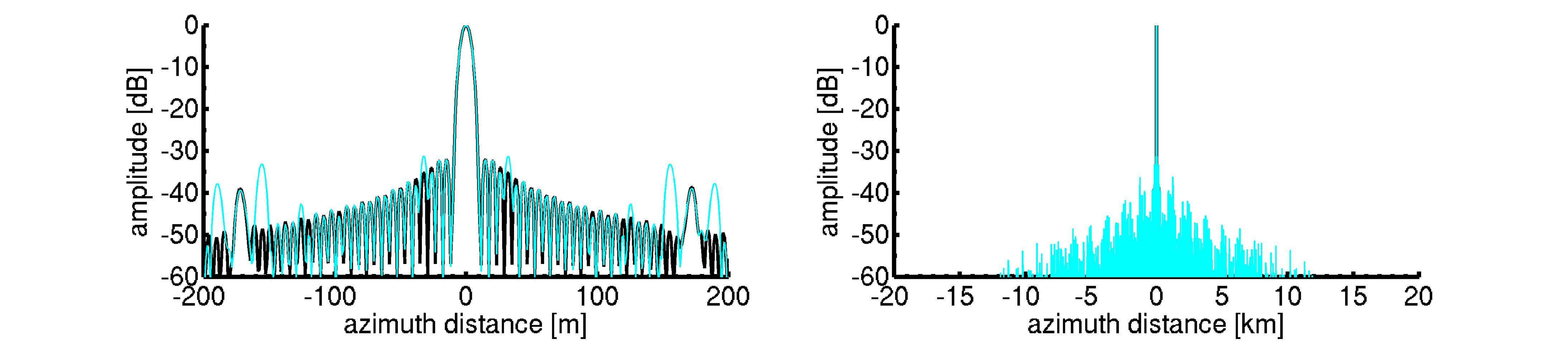}}\hspace*{-0.4in}
  \subfigure[POLYPHASE with elaborate PRI variation.]{%
    \includegraphics[width=0.53\linewidth]{%
      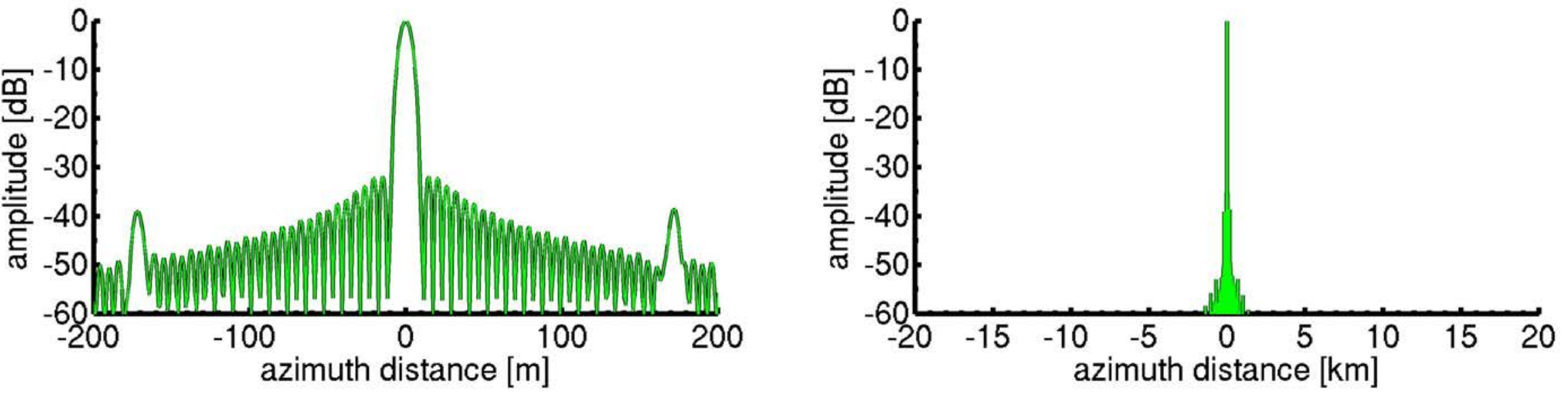}}
  \end{center}
  \vv\vx
  \caption{\scriptsize{Scenario~II (scatterers at $\{-175, 0, +175\}$ (m)): Comparison of BLUI and POLYPHASE using the azimuth IPRs corresponding to the 3 different PRI variations in Fig.~\ref{fig:PRISequences}. Each IPR is depicted  over $400$ (m) and $40$ (km) azimuth scales. The underlying black curves represent the reference IPR of a SAR system with a constant PRI  equal to the mean PRI $PRI_{\tx{mean}}=0.385$ (ms).}}
  \label{Fig:Performance_scenario_2}
\end{figure*}

\bi{Scenario~II (Scatterers at $\{-175, 0, +175\}$ (m)).}
Fig.~\ref{Fig:Performance_scenario_2} shows the azimuth IPRs generated by BLUI and POLYPHASE for the 3 periodic PRI variations in Fig.~\ref{fig:PRISequences} with scatterers at $\{-175, 0, +175\}$ (m). The underlying black curves represent the reference IPR of a SAR system with a constant PRI equal to the mean PRI $PRI_{\tx{mean}}=0.385$ (ms).

Figs~\ref{Fig:Performance} and \ref{Fig:Performance_scenario_2}, and Table~\ref{tab:ISLR_PSLR_Comparison} which compares the two schemes relative to the response of a constant PRI, show that POLYPHASE achieves nearly perfect reconstruction with all 3 of the PRI variations. 
\begin{table}[htpb]
  \centering
  \caption{Integrated Side Lobe Ratio (ISLR) and Peak Side Lobe Ratio (PSLR) for the IPRs in Figs~\ref{Fig:Performance}-\ref{Fig:Performance_scenario_2} for Scenarios I and II}
  \vv
  \scriptsize
  \renewcommand{\arraystretch}{1.1}\addtolength{\tabcolsep}{-3pt}
  \begin{tabular}{ll rr rr rr} 
    \hline
    \hspace*{0in}
      & \hfil\tb{PRI}
      & \mcol{2}{c}{\tb{Reference}}
      & \mcol{2}{c}{\tb{BLUI}}   
      & \mcol{2}{c}{\tb{POLYPHASE}} \\
    \hspace*{0in}
      & \hfil\tb{Variation}
      & \tb{I}\hfil & \tb{II}\hfil 
      & \tb{I}\hfil & \tb{II}\hfil 
      & \tb{I}\hfil & \tb{II}\hfil \\
    \hline\hline
    \tb{ISLR}
      & \tb{Slow} 
      & $-18.30$ & $-18.27$ 
      & $-17.98$ & $-17.94$ 
      & $-18.30$ & $-18.27$ \\ 
    \tb{(dB):}
      & \tb{Fast} 
      & $-18.30$ & $-18.27$ 
      & $-18.23$ & $-18.17$ 
      & $-18.30$ & $-18.27$ \\ 
    \hspace*{0in}
      & \tb{Elaborate} 
      & $-18.30$ & $-18.27$ 
      & $-17.62$ & $-17.66$ 
      & $-18.30$ & $-18.27$ \\
    \hline\hline
    \tb{PSLR}
      & \tb{Slow} 
      & $-32.11$ & $-31.93$ 
      & $-32.04$ & $-31.84$ 
      & $-32.11$ & $-31.93$ \\ 
    \tb{(dB):}
      & \tb{Fast} 
      & $-32.11$ & $-31.93$   
      & $-32.01$ & $-31.89$ 
      & $-32.11$ & $-31.93$ \\ 
    \hspace*{0in}
      & \tb{Elaborate} 
      & $-32.11$ & $-31.93$ 
      & $-31.10$ & $-31.29$ 
      & $-32.09$ & $-31.93$ \\
    \hline
    \end{tabular}
  \label{tab:ISLR_PSLR_Comparison}
\end{table}


\subsection{Real Data}


\begin{figure}[htpb]
  \begin{center}
  \subfigure[PRI sequence of Dataset~A.]{%
    \includegraphics[width=0.48\linewidth]{%
      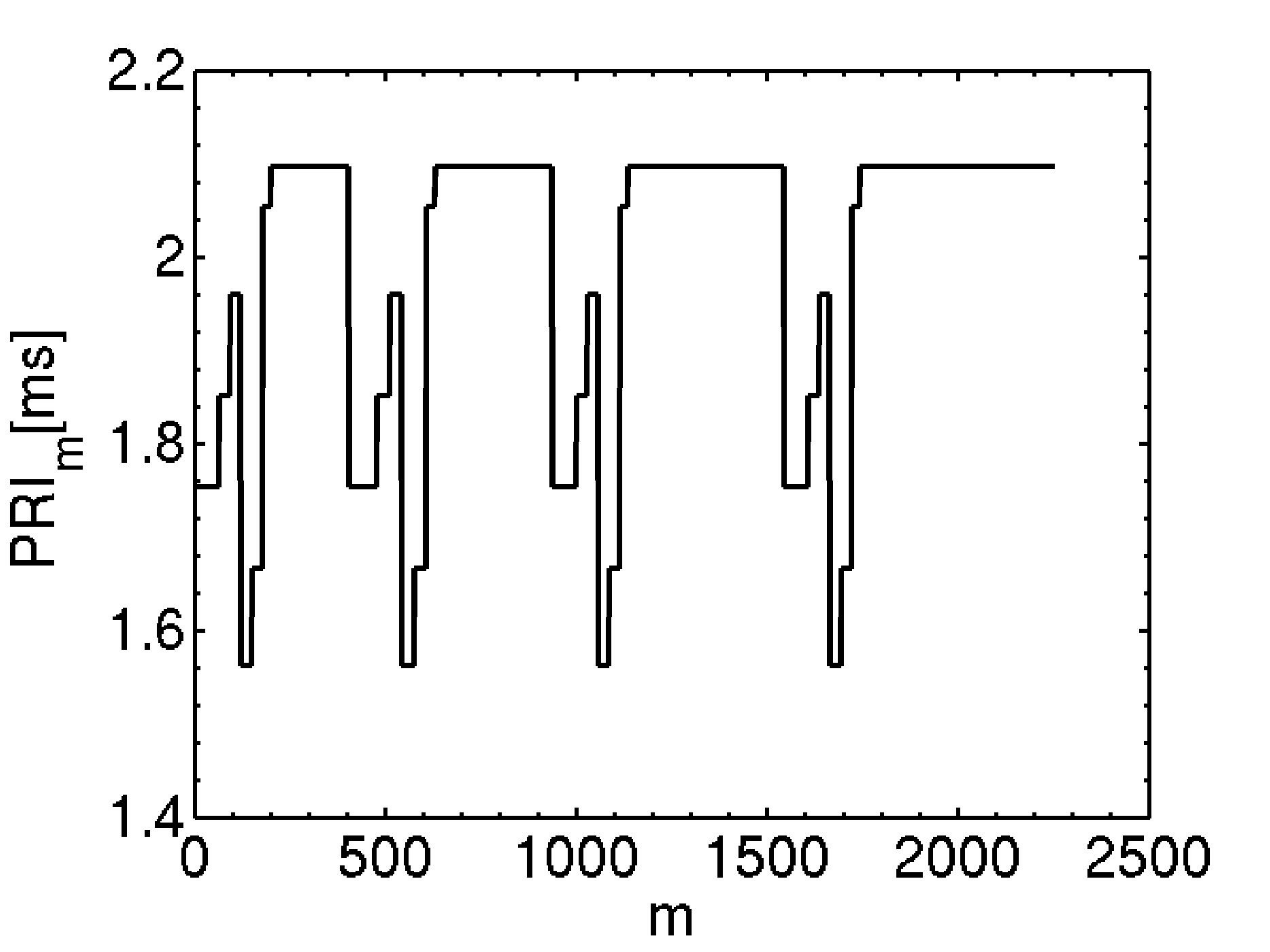}}
  \subfigure[PRI sequence of Dataset~B.]{%
    \includegraphics[width=0.48\linewidth]{%
      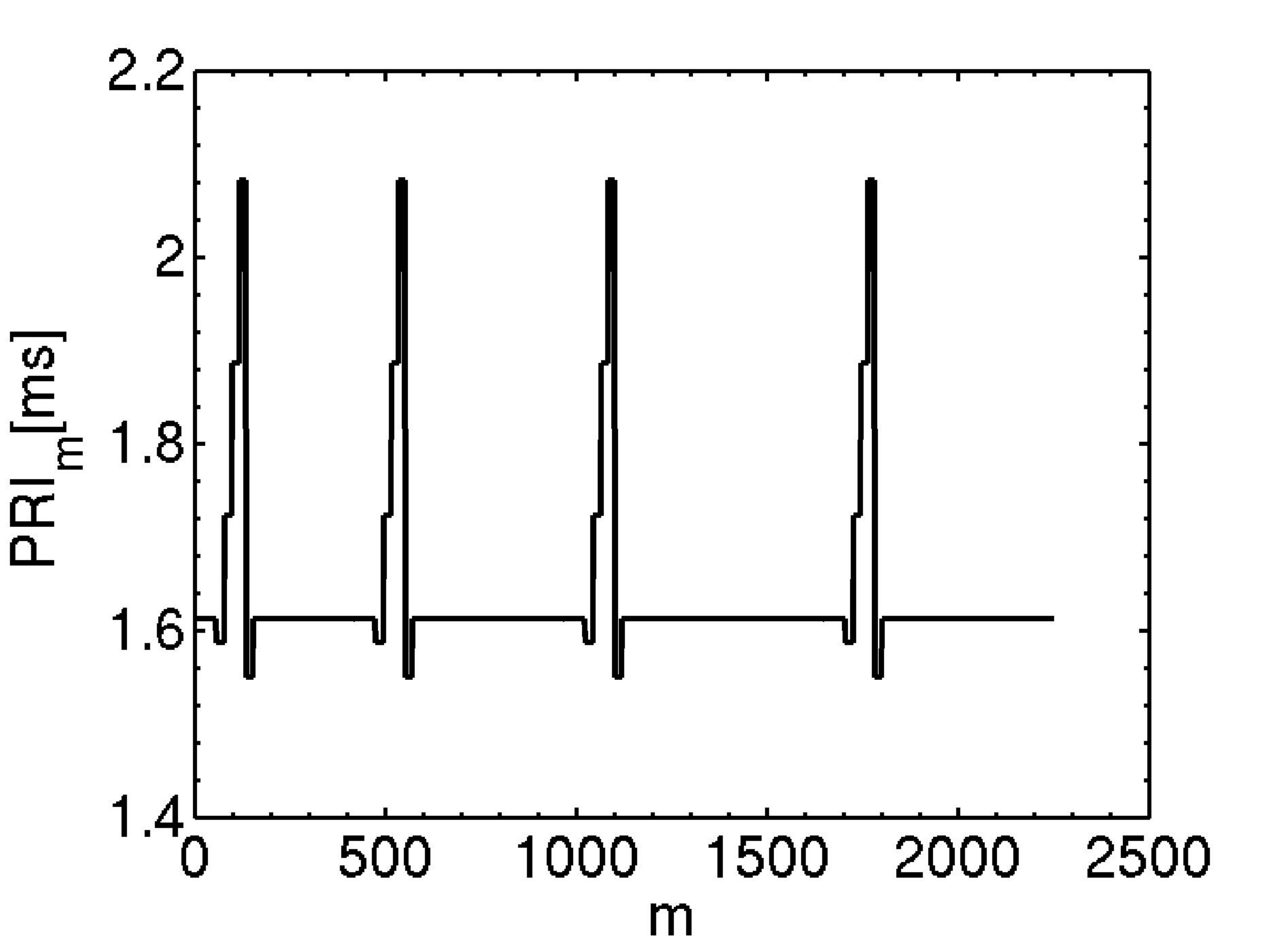}}
  \end{center}
  \vv\vx
  \caption{\scriptsize{Periodic PRI sequences associated with Dataset~A and Dataset~B. The pulse width is $66.7$ ($\mu$s). 
    (a)~Dataset~A: $PRI_{\tx{mean}}=1.984$ (ms), $PRI_{\max}=2.098$ (ms), $PRI_{\min}=1.562$ (ms), $R_0=135$ (km). 
    (b)~Dataset~B: $PRI_{\tx{mean}}=1.637$ (ms), $PRI_{\max}=2.083$ (ms), $PRI_{\min}=1.550$ (ms), $R_0=118$ (km).}}
  \label{fig:PRISequences_RealData}
\end{figure}

\bi{Data Sets.} 
We employ Dataset~A and Dataset~B which had been acquired with stretch waveforms (using the ``deramp-on-receive'' technique) in SAR's spotlight mode, with the PRI slaved to a primary mode possessing the non-uniform PRI variations in Fig.~\ref{fig:PRISequences_RealData}(a) and \ref{fig:PRISequences_RealData}(b), respectively. Table~\ref{tab:notation12} gives the relevant acquisition and processing parameters. The azimuth processed bandwidth is set to $PBW\approx (2/3)PRF_{\tx{out}}$ (i.e., $N_{\tx{save}}=(2/3)\,N_{FFT}$). An azimuth Taylor window (with $\bar{n}=6$, $\tx{sidelobe}=-35$ (dB)) was used to get an azimuth resolution of $0.75$ (m) per resolution cell \cite{Carrara1995}. The azimuth antenna pattern was not compensated for because the antenna pattern as seen by a target is constant in spotlight mode \cite{barbarossa1992detection}. 

\begin{table}[htpb]
  \centering
  \caption{Acquisition and Processing Parameters Corresponding to Datasets~A and B}
  \vv
  \scriptsize
  \renewcommand{\arraystretch}{1.1}\addtolength{\tabcolsep}{-3.5pt}
  \begin{tabular}{c rr rrr rrr} 
    \hline
    \mcol{9}{l}{%
      \tb{Common Acquisition Parameters:}
      $B_{chirp}=240$ (MHz); $\Delta\beta=1.6$ (deg);} \\
    \mcol{9}{l}{%
       $\lambda_c=0.031$ (m); platform height $=12.8$ (km); planar antenna length $=1.45$ (m).} \\
    \mcol{9}{l}{%
      \tb{Common Processing Parameters:} 
      $K_{cr}=1.2$; $L=64$; $N_{FFT}=16384$;} \\
    \mcol{9}{l}{%
       $p_d=1.5$; $Y_{\tx{out}}=3360$ (m).} \\ 
    \hline
    \tb{Data-}
      & $D$ & $R_0$ 
      & $R\Delta\beta$ & $\beta$ & $V_p$ 
      & $PBW$ & $PRF_{\tx{out}}$ & $X_{\tx{out}}$ \\
    \tb{set}
      & (m) & (m)
      & (m) & (deg) & (m/s)
      & (Hz) & (Hz) & (m) \\
    \hline
    \tb{A} 
      & $3395.29$ & $118320$ 
      & $3304.12$ & $13.8$ & $142.68$ 
      & $367$ & $551$ & $7086.94$ \\
    \tb{B}
      & $3821.59$ & $135205$ 
      & $3762.85$ & $10.75$ & $136.59$ 
      & $312$ & $468$ & $7196.15$ \\
    \hline  
    \end{tabular}
  \label{tab:notation12}
\end{table}

\begin{figure}[htpb]
  \begin{center}
  \subfigure[$F(z)$ and $F_{be}(z)$.]{%
    \includegraphics[width=0.48\linewidth]{%
      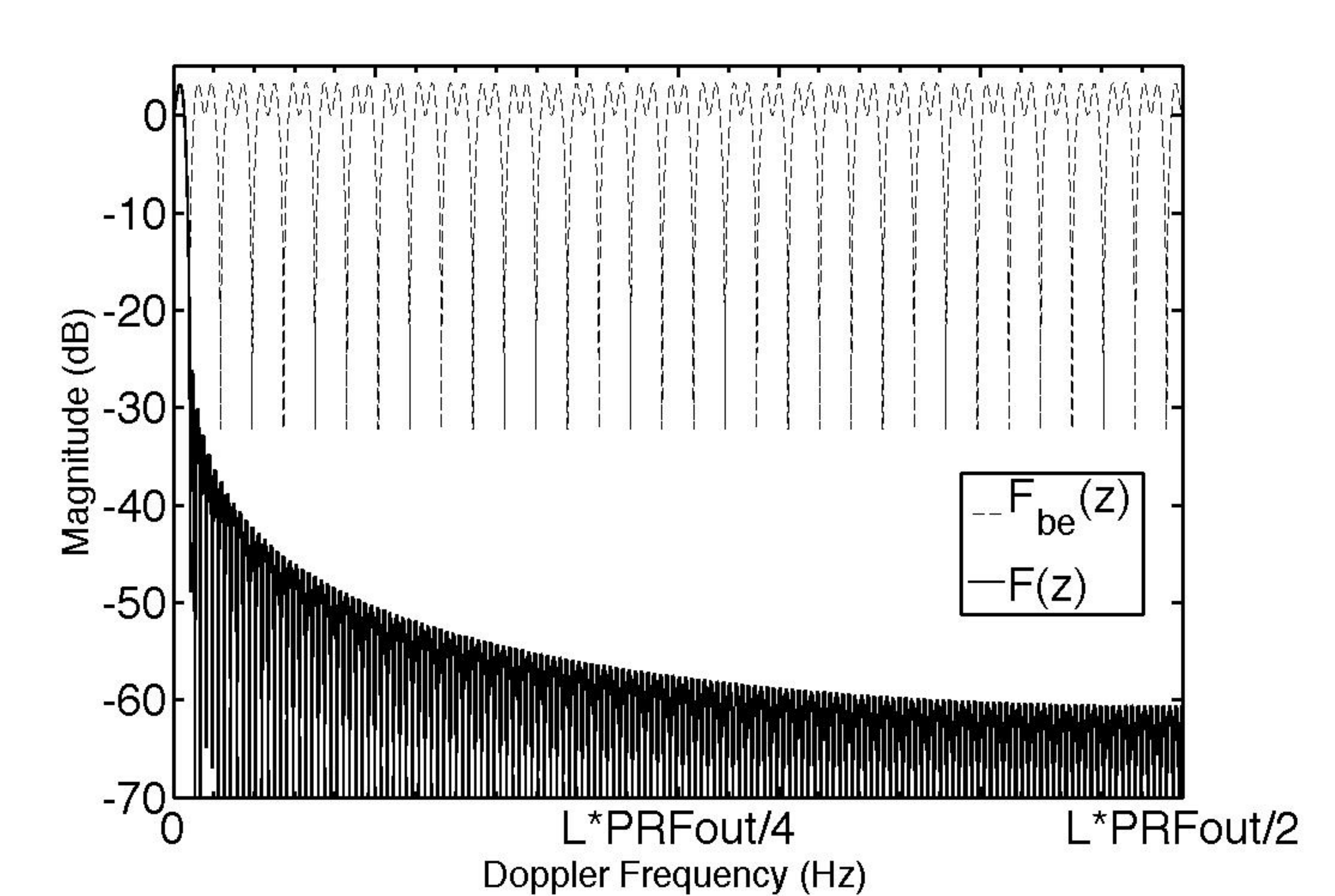}}
  \subfigure[$F_0(z)$.]{%
    \includegraphics[width=0.48\linewidth]{%
      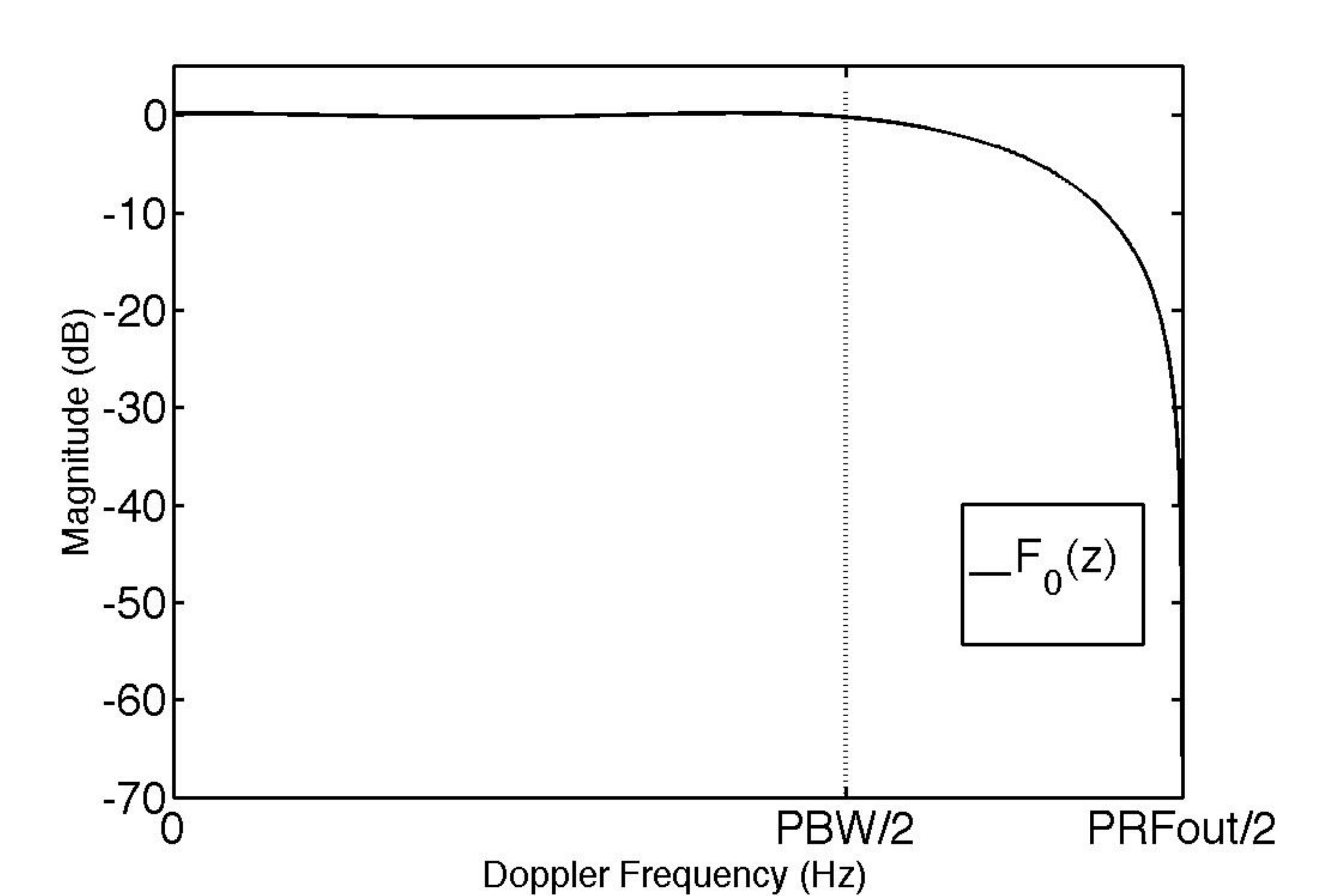}}
  \end{center}
  \vv\vv
  \caption{\scriptsize{Digital filter frequency responses: 
  (a)~Narrowband digital filter $F(z)$ (in \eqref{eq:specs}) and the shaping filter $F_{be}(z)$ (in \eqref{eq:Band Edge Shaping Filter}) both of order $N_{pr}L=(5)(64)=320$. 
  (b)~Polyphase sub-filter $F_0(z)$ (in \eqref{eq:F_l}) of order $N_{pr}=5$.}}
  \label{fig:FilterResponses}
\end{figure}

\bi{Results.}  
Fig.~\ref{fig:FilterResponses}(a) shows the frequency responses of the high-order narrowband digital filter $F(z)$ (in \eqref{eq:specs}) and the shaping filter $F_{be}(z)$ (in \eqref{eq:Band Edge Shaping Filter}); Fig.~\ref{fig:FilterResponses}(b) shows the low-order sub-filter $F_0(z)$ (in \eqref{eq:F_l}).

\begin{figure}[htpb]
  \begin{center}
  \subfigure[Dataset A: non-uniformly sampled along-track.]{%
    \includegraphics[width=0.48\linewidth]{%
      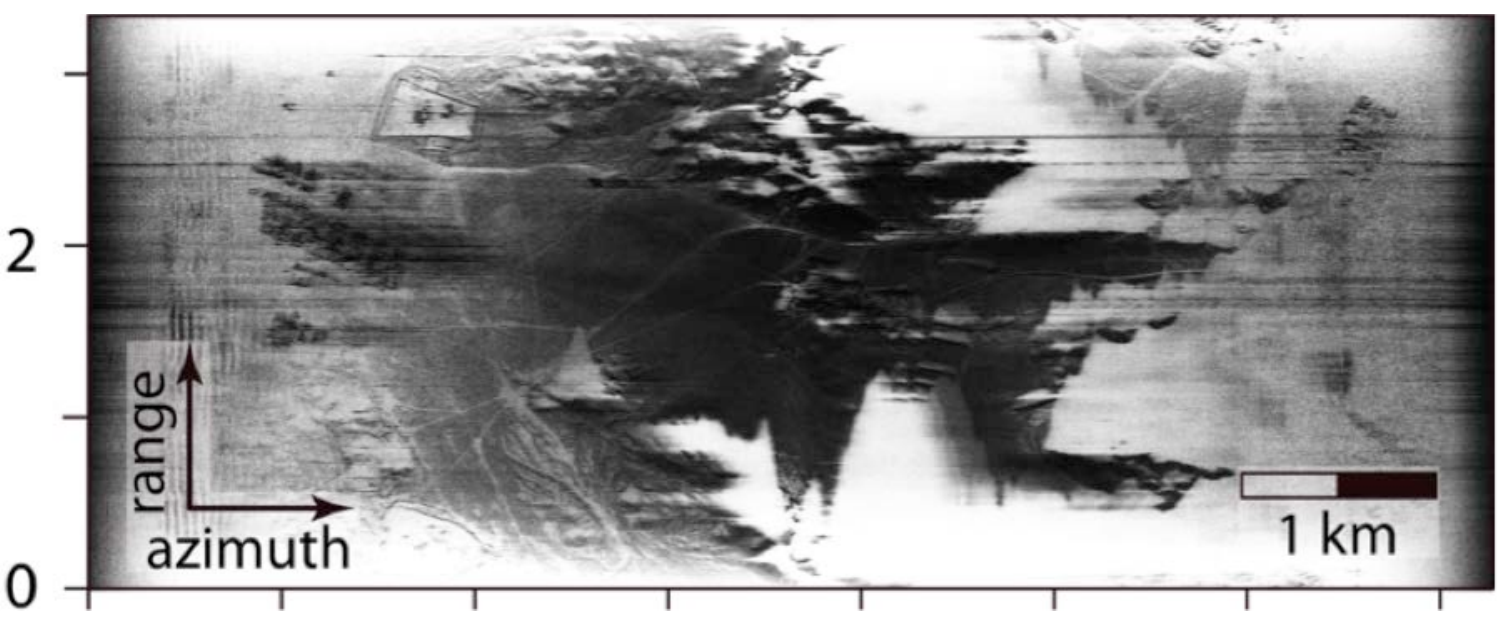}}
  \subfigure[Dataset B: non-uniformly sampled along-track.]{%
    \includegraphics[width=0.48\linewidth]{%
      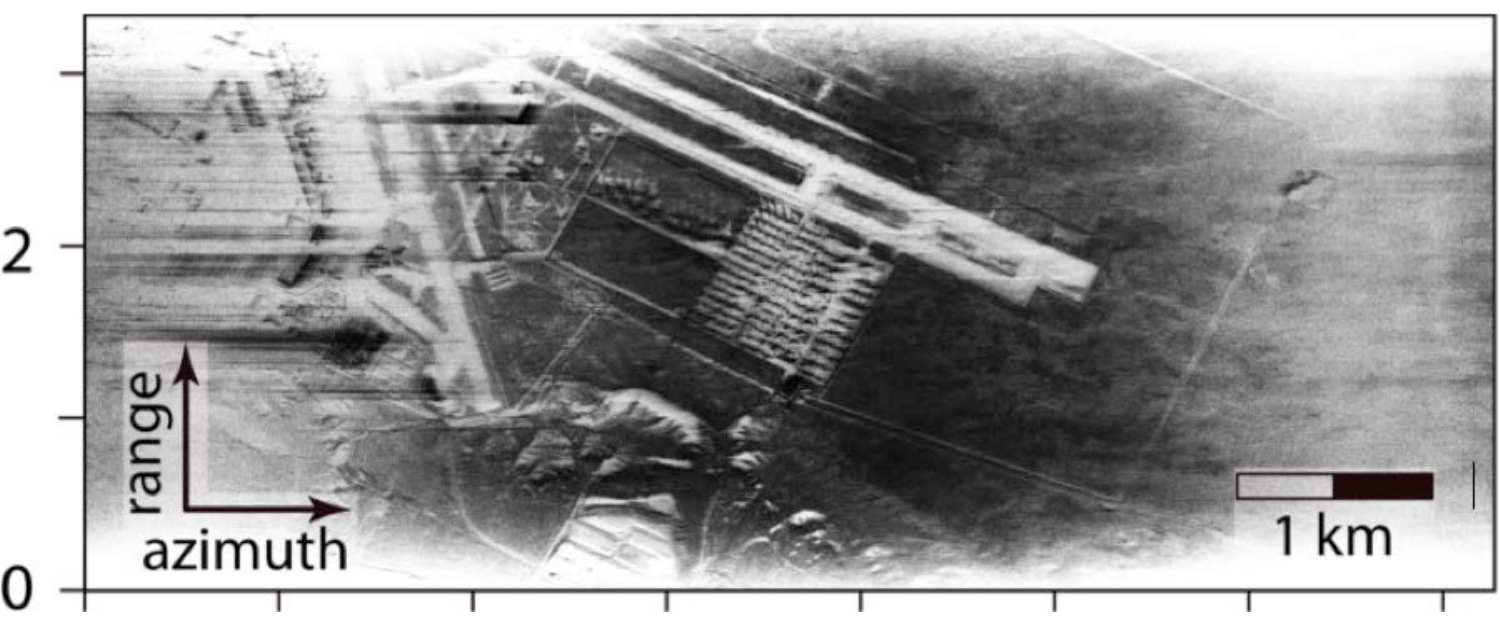}}
  \subfigure[Dataset A: uniformly resampled along-track.]{%
    \includegraphics[width=0.48\linewidth]{%
      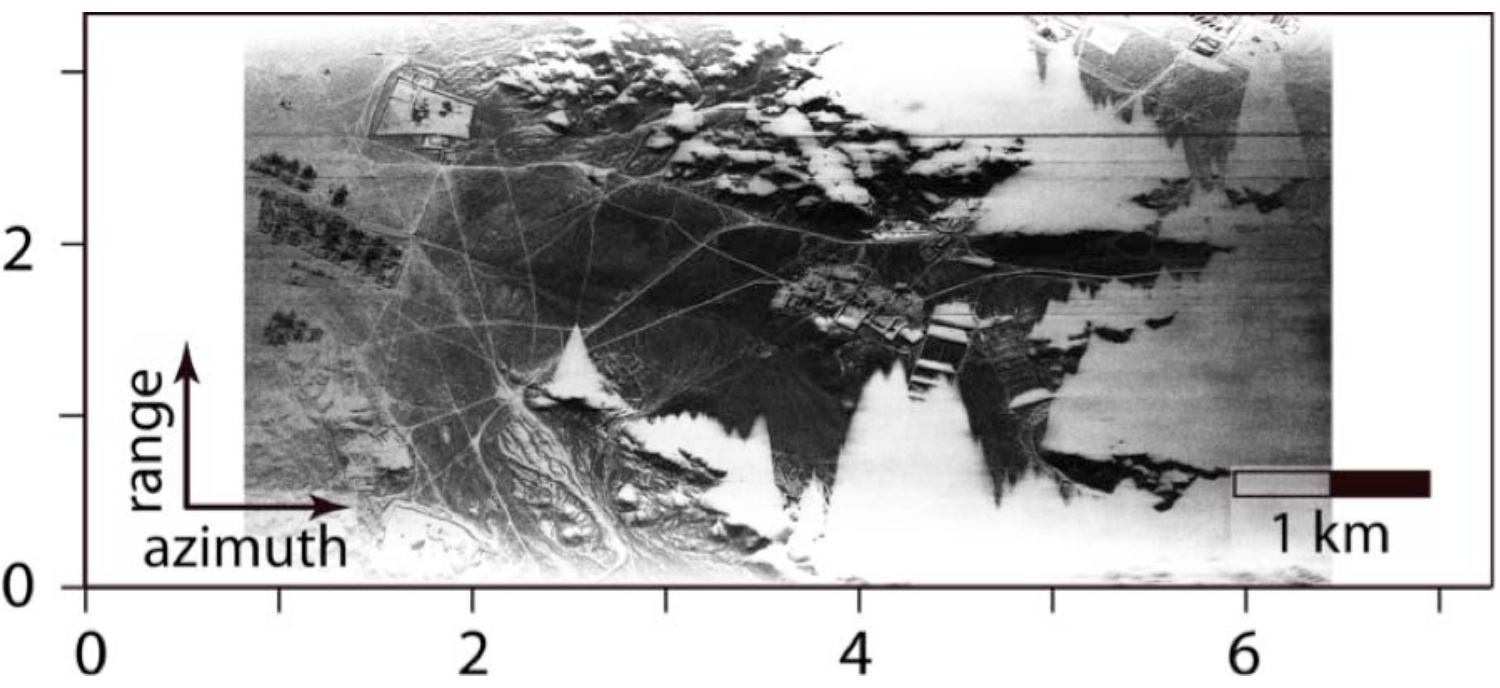}}
  \subfigure[Dataset B: uniformly resampled along-track.]{%
    \includegraphics[width=0.48\linewidth]{%
      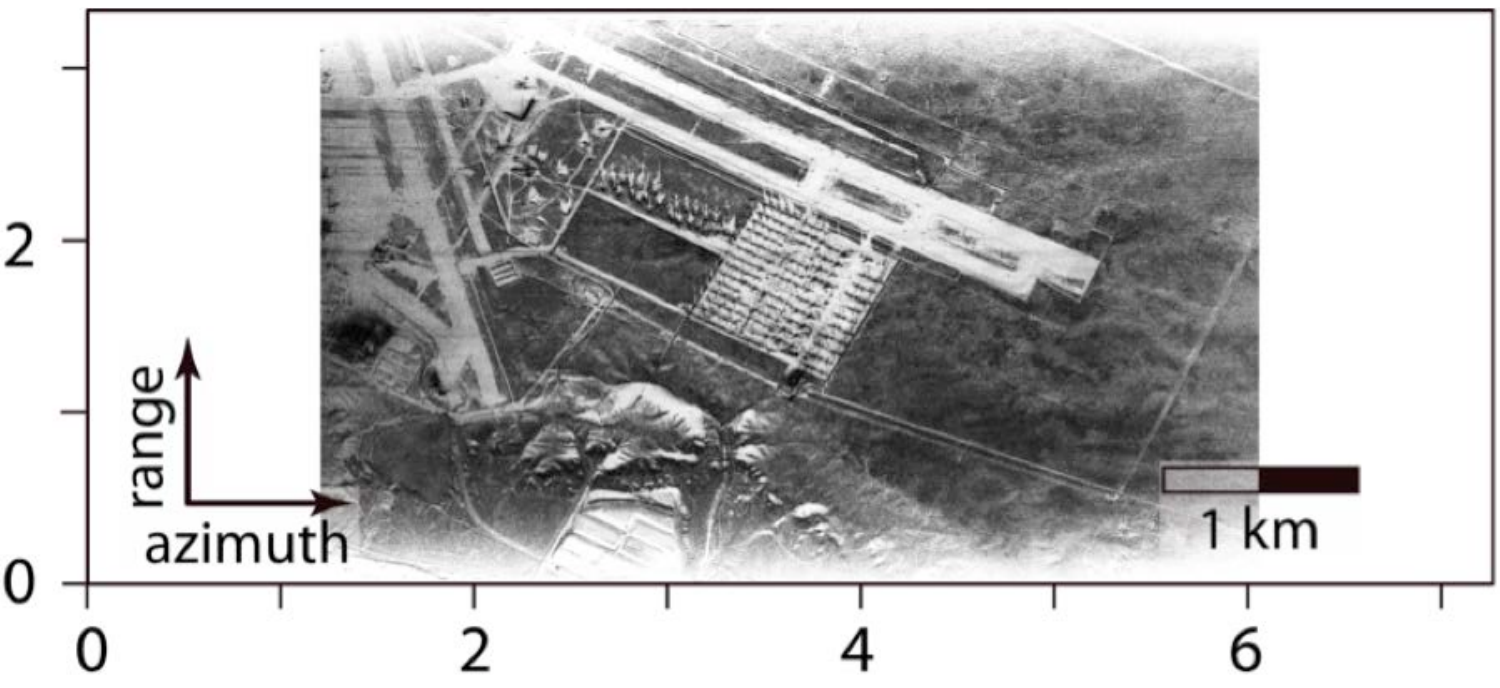}}
  \end{center}
  \vv\vx
  \caption{\scriptsize{Application of POLYPHASE to Datasets~A and B. 
    (a)~Dataset A: non-uniformly sampled along-track formed image; 
    (b)~Dataset B: non-uniformly sampled along-track formed image; 
    (c)~Dataset A: uniformly resampled along-track formed and passband cropped image; 
    (d)~Dataset B: uniformly resampled along-track formed and passband cropped image.}}
  \label{fig:Dataset12}
\end{figure}

Fig.~\ref{fig:Dataset12} shows the results. Figs~\ref{fig:Dataset12}(a) and (c)  refer to Dataset~A: Fig.~\ref{fig:Dataset12}(a) shows the image formed with the spatial FT taken on data non-uniformly sampled along-track, i.e., the resampler block in Fig.~\ref{Fig:Input_Center} is absent; Fig.~\ref{fig:Dataset12}(c) shows the image formed after taking the discrete spatial FT on data that has undergone the resampling system in the spatial slow-time domain, i.e., the resampler block in Fig.~\ref{Fig:Input_Center} in operational. Similarly, Figs~\ref{fig:Dataset12}(b) and (d) refer to Dataset~B: Fig.~\ref{fig:Dataset12}(b) shows the image formed with the spatial FT taken on data non-uniformly sampled along-track; Fig.~\ref{fig:Dataset12}(d) shows the image formed after taking the discrete spatial FT on data that has undergone the resampling system in the spatial slow-time domain. Figs~\ref{fig:Dataset12}(c) and (d) images have been cropped in azimuth to the processing bandwidth $PBW=(2/3)\,PRF_{\tx{out}}$ (or, $N_{\tx{save}}=(2/3)\,N_{FFT}$ pixels). 

Table~\ref{tab:ISLR_PSLR_IMP_Real_Data_Comparison} compares BLUI and POLYPHASE in terms of ISLR and PSLR measurements (relative to the original non-uniformly sampled data).

\begin{table}[htpb]
  \centering
  \caption{Performance Comparison of BLUI and POLYPHASE}
  \vv
  \scriptsize
  \renewcommand{\arraystretch}{1.1}\addtolength{\tabcolsep}{-1.5pt}
  \begin{tabular}{rr rr rr} 
    \hline
    \mcol{2}{c}{\tb{Original (Non-Uniformly}}
      & \mcol{2}{c}{\tb{BLUI}}   
      & \mcol{2}{c}{\tb{POLYPHASE}} \\
    \mcol{2}{c}{\tb{Sampled Data)}} \\
    \tb{Dataset~A} & \tb{Dataset~B} 
      & \tb{Dataset~A} & \tb{Dataset~B} 
      & \tb{Dataset~A} & \tb{Dataset~B} \\
    \hline\hline
    \mcol{6}{l}{\tb{ISLR (dB):}} \\
    $-2.39$ & $1.09$ 
      & $-10.12$ & $-1.95$ 
      & $-10.15$ & $-2.00$ \\ 
    \hline
    \mcol{6}{l}{\tb{PSLR (dB):}} \\
    $-2.0$ & $-2.2$   
      & $-21.0$ & $-9.2$ 
      & $-21.1$ & $-9.3$ \\ 
    \hline
    \mcol{6}{l}{\tb{IPR (m):} (8 scatterers)} \\
    $1.51$, $1.64$ & $1.48$, $1.52$
      & $0.87$, $0.86$ & $0.90$, $0.93$
      & $0.86$, $0.84$ & $0.88$, $0.92$ \\
    $1.78$, $1.89$ & $1.45$, $1.31$
      & $1.07$, $1.21$ & $0.82$, $0.84$
      & $1.05$, $1.20$ & $0.82$, $0.83$ \\
    $0.92$, $0.98$ & $1.23$, $0.88$
      & $0.75$, $0.75$ & $0.78$, $0.73$
      & $0.72$, $0.74$ & $0.76$, $0.72$ \\
    $0.89$, $0.82$ & $0.91$, $0.85$
      & $0.73$, $0.72$ & $0.75$, $0.73$
      & $0.73$, $0.71$ & $0.74$, $0.73$ \\
    \hline
  \end{tabular}
  \label{tab:ISLR_PSLR_IMP_Real_Data_Comparison}
\end{table}

\bi{Resolution Improvement.} 
To quantify the improvement in resolution offered by POLYPHASE, we used IPR measurements and measuring how resolvable point scatterers are in azimuth \cite{russ2011image}. In particular, we employed a quadratic fit of the log magnitude of pixels adjacent to the peak of the main lobe response of point scatterers, and then recorded the $-3$ (dB) width as an indication of resolution. Table~\ref{tab:ISLR_PSLR_IMP_Real_Data_Comparison} indicates the average improvement corresponding to 8 point scatterers. 


\section{Discussion}
\label{sec:Discussion}


\subsection{Computational Complexity}


We use the common practice of counting the number of \emph{flops} (i.e., the number of real additions and multiplications) to compare the computational burdens of BLUI and POLYPHASE \cite{Axelsson1996, Johnson2007ToSP}. Note that BLUI is \emph{output-based} (i.e., it takes in a vector of input samples to compute an output sample); POLYPHASE is \emph{input-based} (i.e., it takes an input sample and updates a vector of output samples). 

\bi{POLYPHASE.} 
Each complex-valued input sample $r(n)\in\mbb{C}$ gets operated on by $N_{pr}$ real-valued taps of only one polyphase component to update $N_{pr}$ output samples (see Fig.~\ref{fig:PolyphaseImplementation}), entailing $4N_{pr}$ flops ($2N_{pr}$ real multiplications and additions each). An $N_D$-length output vector (to represent the synthetic aperture length $D$) needs about $PRI_{\tx{out}}/\tx{mean}[PRI_{\tx{in}}]\cdot N_D$ input samples. So, \emph{each} output sample requires about $4N_{pr}PRI_{\tx{out}}/\tx{mean}[PRI_{\tx{in}}]$ (flops).

\bi{BLUI.} 
Here, non-uniformly sampled input SAR data are interpolated, Doppler filtered, and then decimated to produce the uniformly sampled output signal $y(\cdot)$ \cite{Villano2014ToGRS, Villano2014IRS, Villano2015APSAR}. 

\emph{Interpolation Stage.} 
To compute the interpolated sample $r(n)$, BLUI uses $Q$ non-uniformly sampled input samples of $s'(\cdot)$ located at $\mc{Q}_n\equiv\{n_1, \ldots, n_Q\}$ and falling within the interval $[-L_{\tx{RA}},+L_{\tx{RA}}]$ on either side of $n$. Here, $L_{\tx{RA}}$ is the real aperture length, $V_p$ (m/s) is the platform velocity, and $Q=\lfloor\left(2\,L_{\tx{RA}}/PRI_{\tx{in}}V_p\right)\rfloor$. The interpolated value is 
\begin{equation}
  r(n)
    =\mb{b}(\mc{Q}_n^+)^T\widehat{\mb{s}}'(\mc{Q}_n).
\end{equation}
Here, $\widehat{\mb{s}}'(\mc{Q}_n)=[s'(n_1), \ldots, s'(n_Q)]^T\in\mbb{C}^{Q\times 1}$ is the input data value `segment' and $\mb{b}(\mc{Q}_n^+)=\mb{G}(\mc{Q}_n)^{-1}\pmb{\varrho}(\mc{Q}_n^+)\in\mbb{R}^{Q\times 1}$, where $\mb{G}(\mc{Q}_n)\in\mbb{R}^{Q\times Q}$ is a certain symmetric matrix (with equal diagonal entries and $\pmb{\varrho}(\mc{Q}_n^+)\in\mbb{R}^{Q\times 1}$ is a function of both $n$ and $\mc{Q}(n)$, i.e., $\mc{Q}_n^+=\{n, \mc{Q}(n)\}$. 

\emph{Filtering and Decimation Stages.} 
Interpolated data $r(\cdot)$ are Doppler filtered using a $N_{\tx{BLUI}}$-tap filter with $N_{\tx{BLUI}}=25$ or $17$. A $9$-tap low-order Capon beamformer provides acceptable passband, but introduces significant passband attenuation requiring additional compensation during processing \cite{capon1969high, Villano2014IRS}. 

\emph{Joint BLUI.} 
To avoid the heavy a computational burden imposed by the above 3-stage implementation, a \emph{joint BLUI scheme} wherein the matrix operations are carried out on-ground and the 3 stages are jointly conducted on-board is suggested in \cite{Villano2014ToGRS, Villano2015APSAR}, without further details of such a scheme. 

To compare the computational complexity of BLUI and POLYPHASE, we now develop this joint BLUI scheme. Note that the interpolated sequence $r(\cdot)$ is first sent through an order-$N_{BLUI}$ digital filter with coefficient vector $\mb{h}=[h_0, \ldots, h_{N_{BLUI}}]^T$ and then decimated by $L_{BLUI}$ to get 
\begin{equation}
  y(n)
    =\!\!\sum_{k=0}^{N_{BLUI}}\!\! 
     h_k\,r(n'-k)
    =\!\!\sum_{k=0}^{N_{BLUI}}\!\!
     h_k\,\mb{b}(\mc{Q}_{n'-k}^+)^T\,\widehat{\mb{s}}'(\mc{Q}_{n'-k}),
  \label{eq:r}
\end{equation}
where, for notational convenience, we use $n'=nL_{BLUI}$.

\emph{(a)~Arbitrary Input PRF Variation.} 
Entries of $\widehat{\mb{s}}'(\cdot)$, $\mb{G}(\cdot)$, $\pmb{\varrho}(\cdot)$, and hence $\mb{b}(\cdot)$, are functions of $\mc{Q}_n$ and the location $n$ where the signal value is being interpolated. So, one must freshly generate $\mb{G}(\mc{Q}_{n'})$, $\pmb{\varrho}(\mc{Q}_{n'}^+)$, and $\mb{b}(\mc{Q}_{n'}^+)$ on-ground, and up-link the latter, for \emph{each} $n$. Regarding on-ground computations, $\pmb{\varrho}(\mc{Q}_{n'}^+)$ and $\mb{G}(\mc{Q}_{n'})$ require $\tx{F}_R\,Q$ (flops) and $\tx{F}_R\,(Q^2-Q+2)/2$ (flops), respectively, where $\tx{F}_R$ is the flop count for computing each entry of the input azimuth signal's autocorrelation function $R_u(\xi)$ \cite{Villano2014ToGRS}; and $\mb{b}(\mc{Q}_{n'}^+)=\mb{G}(n)^{-1}\pmb{\varrho}(n)$ requires $2Q^3/3+3Q^2/2-7Q/6$ (flops) with  Gaussian elimination (and backward substitution) \cite{Bunch1971JNA, Axelsson1996}. Regarding on-board computing of $y(n)$, computing $r(n'-k)$, multiplying by $h_k$, and then adding the $N_{BLUI}+1$ terms in \eqref{eq:r} requires $4Q(N_{BLUI}+1)-1$ (flops). 

When input PRF variation is arbitrary, Table~\ref{tab:compare} compares the flop counts of both schemes for the synthetic data cases in \cite{Villano2014ToGRS} corresponding to the fast PRI and elaborate PRI sequences (results for slow and fast PRI cases are identical). Clearly, POLYPHASE offers a significant computational advantage ($10{\sim}15$ times in on-board flops, and much more in total flops).

\begin{table*}[htpb]
  \centering
  \caption{\# of Flops Required to Compute (on a \emph{Per-Output} Basis)}
  \vv
  \scriptsize
  \renewcommand{\arraystretch}{1.1}\addtolength{\tabcolsep}{0pt}
  \begin{tabular}{l l p{1.20in} rr r} 
    \hline
    \hfil\tb{Scheme}
      & \hfil\tb{Operation}
      & \hfil\tb{Flop Count Estimate}
      & \tb{Fast PRI}\hfil
      & \tb{Elaborate PRI}\hfil 
      & \tb{Staggered} \\
    {}
      & 
      & 
      & \tb{Variation in \cite{Villano2014ToGRS}}\hfil
      & \tb{Variation in \cite{Villano2014ToGRS}}\hfil 
      & \tb{System in \cite{Villano2015APSAR}} \\
    \hline\hline 
    \emph{Common Parameters}
      & \mcol{4}{l}{%
          BLUI: $L_{\tx{RA}}=10$ (m), $L_{\tx{BLUI}}=3$, $N_{\tx{BLUI}}=9$, 
          $V_p=7,473$ (m/s), $\tx{F}_R\approx 24$; 
          POLYPHASE: $N_{pr}=5$.} \\
    \emph{Other Parameters}
      & $PRI_{\tx{in}}$ (ms)
      & 
      & $[0.349, 0.421]$ \hfil
      & $[0.309, 0.461]$ \hfil 
      & --- \\ 
    {}
      & \emph{$Q$}
      &
      & $6{\sim}7$\hfil 
      & $5{\sim}8$\hfil 
      & $5{\sim}8$\hfil \\
    {}
      & $\tx{mean}[PRI_{in}]$ (ms)
      &  
      & $0.385$
      & $0.385$
      & $0.37037$ \\
    \hline\hline     
    \tb{BLUI}
      & $PRI_{out}$ (ms)
      &
      & (a)~$0.417$
      & (a)~$0.417$ 
      & \\
    \tb{(Arbitrary)} 
      & Up/Down-link floats
      & $2Q$
      & $12{\sim}14$ 
      & $10{\sim}16$ 
      & \\
    {}
      & \tb{ON-BOARD flops}
      & $4Q(N_{\tx{BLUI}}{+}1)+2N_{BLUI}$
      & $\mb{258{\sim}298}$ 
      & $\mb{218{\sim}338}$ 
      & \\
    {}
      & On-ground flops
      & $2Q^3/3+Q^2(\tx{F}_R{+}3)/2+Q(\tx{F}_R{-}7/3)/2+\tx{F}_R$
      & $\mb{719{\sim}990}$ 
      & $\mb{499{\sim}1{,}316}$ 
      & \\
    \hline
    \tb{BLUI}
      & $PRI_{out}$ (ms)
      &
      & (b)~$0.770$
      & (b)~$0.770$
      & $1.11111$ \\
    \tb{(Periodic)}
      & $\Psi$ (from \eqref{eq:Psi})
      &
      & $1$
      & $1$
      & $1$ \\
    {}
      & Up/Down-link floats
      & $N_{BLUI}\Psi+Q$
      & $15{\sim}16$
      & $14{\sim}17$ 
      & $14{\sim}17$ \\
    {}
      & \tb{ON-BOARD flops}
      & $4(N_{BLUI}\Psi{+}Q)-2$
      & $\mb{58{\sim}62}$
      & $\mb{54{\sim}66}$ 
      & $\mb{54{\sim}66}$ \\
    {}
      & On-ground flops
      & Same as for arbitrary case;
      &
      & 
      & \\
    {}
      &
      & but computed only $R$ times
      &
      & \\
    \hline\hline
    \tb{POLYPHASE}
      & \tb{ON-BOARD flops}
      & $4N_{pr}PRI_{\tx{out}}/\tx{mean}[PRI_{\tx{in}}]$
      & $\mb{22}$
      & $\mb{22}$ 
      & $\mb{60}$ \\
    \mcol{2}{l}{\tb{(Arbitrary, including Periodic)}}
      &&&& \\
    \hline
    \mcol{5}{l}{%
      (a)~We use $PRI_{out}=0.417$ (ms) because the performance comparison  in Section~\ref{sec:SimulatedData} is conducted with the same value.} \\
    \mcol{5}{l}{%
      (b)~$PRI_{out}=0.770$ (ms) is used so that \eqref{eq:periodicity} is satisfied and the computational simplification of the BLUI periodic case can be harnessed.}
    \end{tabular}
  \label{tab:compare}
\end{table*}

\emph{(b)~Periodic Input PRF Variation.} 
Suppose, as in staggered SAR, the input PRI sequence repeats every $T_{PRI}$\,(s), or every $N_{PRI}$ samples. Then, suppose an integer number $R$ of (pre-decimation) output samples `fit' within $T_{PRI}$\,(s), i.e., 
\begin{equation}
  R
    \equiv
     L_{BLUI}T_{PRI}/PRI_{out}
    \in\mbb{N}_+.
  \label{eq:periodicity0}
\end{equation}
Then, BLUI could be made computationally very efficient because $\mc{Q}_n$, and therefore $\mb{G}(\mc{Q}_n)^{-1}$, $\pmb{\varrho}(\mc{Q}_n^+)$, and $\mb{b}(\mc{Q}_n^+)$, also repeat every $R$ samples. In staggered SAR \cite{Villano2015APSAR}, $T_{PRI}=N_{PRI}\tx{mean}[PRI_{in}]$, and \eqref{eq:periodicity0} reduces to 
\begin{equation}
  R
    =L_{BLUI}N_{PRI}\tx{mean}[PRI_{in}]/PRI_{out}
     \in\mbb{N}_+.
  \label{eq:periodicity}
\end{equation}

The work \cite{Villano2015APSAR} in fact satisfy \eqref{eq:periodicity} with $L_{BLUI}=3$ and $\{\tx{mean}[PRI_{in}], PRI_{out}\}=\{0.37037, 1.11111\}$. But, this is not true with the input PRI variations in Fig.~\ref{fig:PRISequences} when the parameters in \cite{Villano2014ToGRS} are used. Instead, one can exploit joint BLUI's  computational advantage by selecting $PRI_{out}=0.770$ (ms) (instead of $PRI_{out}=0.417\,\tx{(ms)}$ used in \cite{Villano2014ToGRS}).

When \eqref{eq:periodicity} is satisfied, instead of \eqref{eq:r}, we may use 
\begin{equation}
  y(n)
    =\sum_{k=0}^{N_{BLUI}} 
     h_k\,\mb{b}(\mc{Q}_{n'-k}^+)_R^T\,\widehat{\mb{s}}'(\mc{Q}_{n'-k}),
  \label{eq:rR}
\end{equation}
where $(\cdot)_R$ denotes the modulo-$R$ operation. So the entries of $\mb{b}(\cdot)$ are reusable and only $R$ of its samples need be computed. 

Moreover, an overlap between consecutive segments $\widehat{\mb{s}}'(\mc{Q}_{n'-k})$ and $\widehat{\mb{s}}'(\mc{Q}_{n'-k+1})$ leads to further computational reduction. For example, suppose the first entry of the segment $\widehat{\mb{s}}'(\mc{Q}_{n'-N_{BLUI}})$ is $s'(1)$ and only $\Psi$ samples (where $1\leq\Psi\leq Q$) are `new' between consecutive segments. In fact, an approximate expression for $\Psi$ is 
\begin{equation}
    \Psi
      \approx
       \min
       \left\{
         \lceil
           {PRI_{out}/L_{BLUI}\tx{mean}[PRI_{in}]}
         \rceil, Q
       \right\}.
  \label{eq:Psi}
\end{equation}
Then, one can show that only $N_{BLUI}\Psi+Q$ samples of the non-uniformly sampled signal $s'(\cdot)$ are needed to compute one output sample. This allows \eqref{eq:rR} to be expressed as
\begin{equation}
  y(n)
    =\sum_{m=1}^{N_{BLUI}\Psi+Q}
     c_m\,s'(m),
\end{equation}
for appropriately chosen coefficients $c_m$ (which are real-valued and  depend on $\mb{h}$ and $\mb{b}(\cdot)$). Assuming the coefficients are computed on-ground and uploaded, each output sample would require about $4(N_{BLUI}\Psi+Q)-2$ (flops).  

When input PRF variation is periodic and \eqref{eq:periodicity} holds true, Table~\ref{tab:compare} compares the flop counts of the two schemes. BLUI  shows a significant improvement when compared with its arbitrary case. However, for the fast PRI and elaborate PRI sequences, POLYPHASE still offers a computational advantage of about $2.5$ times in on-board flops; for the staggered system in \cite{Villano2015APSAR}, on-board flop counts are comparable. What must be emphasized here is that POLYPHASE applies to arbitrary PRI variations. Moreover, it requires neither up/down-linking nor on-ground computation of intermediate variables. The computation of $R_u(\xi)$ may require numerical schemes \cite{Villano2014ToGRS}, which may further exacerbate the BLUI's on-ground effort. Moreover, since $R_u(\xi)=0,\,\forall |\xi|\geq L_{\tx{RA}}/V_p$ \cite{Villano2014ToGRS}, BLUI mandates a lower bound on the minimum real aperture length $L_{\tx{RA}}$. POLYPHASE makes no such demands. 


\subsection{Computational Considerations}

\bi{Convolution Computation.} 
In our implementation of POLYPHASE, we re-index $y'(n)$ to work with $\tilde{y}'(n)=y'(n+(N_{pr}+1)/2)$, because the associated inequalities are more symmetric, e.g., corresponding to \eqref{eq:ell2n} we get
\begin{equation}
  \left\lfloor{%
    \frac{\ell-1}{L}}
  \right\rfloor
    -\left(
       \frac{N_{pr}-1}{2}
     \right)
    \leq n
    \leq
     \left\lfloor{%
       \frac{\ell}{L}}
     \right\rfloor
       +\left(
          \frac{N_{pr}-1}{2}
        \right).
  \label{eq:ell2n_s}
\end{equation}

The entries in the output signal vector $\widetilde{\tb{y}}'(N_1{:}N_2)=[\tilde{y}'(N_1), \tilde{y}'(N_1+1), \ldots, \tilde{y}'(N_2)]^T$ are simultaneously and efficiently computed via an \emph{input-centered convolution} scheme \cite{Carrara1995}. See Algorithm~\ref{alg:1}. 

\begin{algorithm}
\fontsize{10}{10}\selectfont
  \begin{algorithmic}
    \STATE Initialize: $\widetilde{\tb{y}}'(N_1{:}N_2)=\tb{0}$; 
    \FOR {$r(\ell)\neq 0$, s.t. 
          $\ell\in\mbb{N}$ satisfies \eqref{eq:n2ell_s_max},}
      \STATE $m=\lfloor{(\ell-1)/L}\rfloor+1$;  
      \FOR {$n\in\mbb{N}$, s.t. $n$ satisfies \eqref{eq:ell2n_s},}
        \STATE $\tilde{y}'(n)=\tilde{y}'(n)
                                +f_{mL-\ell}(n+(N_{pr}+1)/2-m)\,r(\ell)$;
      \ENDFOR
    \ENDFOR
  \end{algorithmic}
  \caption{Input-Centered Convolution Algorithm}
  \label{alg:1}
\end{algorithm}

Note that, the computation of $\widetilde{\tb{y}}'(N_1{:}N_2)$ requires all the non-zero input samples $r(\ell)\neq 0,\,\ell\in\mbb{N}$, s.t. 
\begin{equation}
  \left(
    N_1-\frac{N_{pr}-1}{2}
  \right)L
    \leq\ell
    \leq
     \left(
       N_2+\frac{N_{pr}+1}{2}
     \right)L.
  \label{eq:n2ell_s_max}
\end{equation}
The output $\widetilde{\tb{y}}'(N_1{:}N_2)$ from Algorithm~\ref{alg:1} yields $\tb{y}'(N_1+(N_{pr}+1)/2{:}N_2+(N_{pr}+1)/2)$. The grid alignment procedure (see Section~\ref{subsec:3c}) can be incorporated directly into this computation without having to account for it at an earlier stage. Indeed, as and when a new non-zero signal pulse $s'(k)=s(\nu_r, \alpha_k)$ is received, one can generate $r(n)$ in \eqref{eq:input_rescale} and invoke Algorithm~\ref{alg:1}. In this manner, the computation is carried out `on-the-fly' with no need to buffer the input signal pulses. The implementation waits for non-zero pulses only and processes them one pulse at a time as they are received. 

\bi{Buffer Memory.} 
The implementation occurs in-place, and requires a buffer of size $N_D$ complex floats to store the output, a buffer of ($N_{pr}L+1$) real floats to store the filter coefficients, and one complex float to store the current input temporarily for processing (and later overwrite it with the next input).  

Since $N_D=(N_{FFT}/p_d)\,K_{cr}$ (see \eqref{eq:pdensity}), the parameters $N_{FFT}$, $p_d$, and/or $K_{cr}$ can be used to significantly lower the number of output pulses. However, lowering the FFT size narrows the unambiguously sampled cross-range extent being imaged; lowering $K_{cr}$ broadens the resolution; and increasing $p_d$ increases the spectrum oversampling factor. 

\bi{Real-Time Application.}  
POLYPHASE executes in real-time, i.e., it processes the input pulses one-by-one as they are received with no delay. It handles input pulses even if they are received out-of-order (assuming they are spatially correctly stamped of course). BLUI does not possess this feature. 

\bi{Data Up/Down-Link.} 
With its significantly lower computational burden, \emph{all} processing in  POLYPHASE can be performed on-board with no need for on-ground computations or up/down-linking of intermediate variables. These translate into lower transmit power (an important consideration when using drones or other UAVs), cheaper and lighter communication systems with smaller channel capacities, and faster data transfers with reduced need for retransmission \cite{Proakis2001}.

\bi{Error Analysis.} 
Considering the floor operation associated with the grid alignment process in \eqref{eq:input_rescale}, we may upper bound the error in sample realignment by 
\begin{equation}
  \Delta u_{\tx{error}}
    \leq
     \frac{\Delta u_{\tx{out}}}{L}
     =\frac{1}{L}\,\dfrac{D}{N_D}
     =\frac{D}{L}\,\frac{p_d}{N_{FFT}}\,\frac{1}{K_{cr}}.
  \label{eq: interp error}
\end{equation}
So, $\Delta u_{\tx{error}}$ can be significantly reduced by using larger $N_{FFT}$, which increases the computational load in SAR processing, or larger $L$, which calls for a larger number of sub-filters. But POLYPHASE's computational load remains unaltered because it is a function of $N_{pr}$ (and not $L$). Of course, a larger $L$ may produce a better interpolation and hence a better output.


\subsection{Other Design Considerations} 


\bi{FFT Size.} 
While it is common to select the FFT size for azimuth compression to accommodate a required cross-range extent, we employed a different strategy: We first selected an FFT size $N_{FFT}$ that can be comfortably implemented with the given system resources and used the cross-range extent $X_{\tx{out}}$ which can be accommodated with this $N_{FFT}$ value (see Section~\ref{sec:OutputGrid}). This strategy allows one to partition an image into smaller patches and still operate on the same FFT size. 

\bi{Prototype Digital Filter.} 
Given the cross-range extent associated with the selected FFT size, the FIR prototype filter was designed to pass only a portion (we used $\gamma=N_{\tx{save}}/N_{FFT}=PBW/PRF_{\tx{out}}=2/3$) of the spectrum (see Fig.~\ref{Fig:Y_and_S}). This was necessary to arrive at a low order filter design so that the number of filtering operations is kept in check.
  
\bi{Type of Digital Filter.} 
An FIR digital filter of low order ($N_{pr}=5$) was adequate for our purposes. In addition to the absence of stability issues, the FIR design allowed us to compute only every $L$-th output sample that is  affected by an incoming input sample. This property was critical for implementing the downsampling portion of our system.

\bi{Oversampling of Final Image.} 
Oversampling the final image generated a final image that was more pleasing to the eyes. We used an oversampling factor of $p_d=1.5$ to describe each square radar resolution cell of $\delta_r\times\delta_{cr}$.

\bi{Missing Samples Case.} 
To see how POLYPHASE recovers the signal when additional samples are missing (due to simultaneous Tx/Rx events) in staggered SAR, we randomly removed 10\% of the samples from each PRI sequence in Fig.~\ref{fig:PRISequences}. We used both ISLR and PSLR measures: 
  \tb{(a)}~ISLR: With both the methods in \cite{holm1991performance} and \cite{zenere2012sar}, POLYPHASE consistently outperforms the BLUI by at least by 1.5 (dB) or better when samples are missing; the performance difference is much closer when samples are not missing. 
  \tb{(b)}~PSLR: Here, BLUI was better with the elaborate PRI sequence; POLYPHASE was better with both the slow and fast PRI sequences. 

\begin{table}[htpb]
  \centering
  \caption{Comparison of BLUI and POLYPHASE When 10\% Samples are Missing}
  \vv
  \scriptsize
  \renewcommand{\arraystretch}{1.1}\addtolength{\tabcolsep}{-2.0pt}
  \begin{tabular}{r rrr rrr} 
    \hline
    \tb{Reference}
      & \mcol{3}{c}{\tb{BLUI}}   
      & \mcol{3}{c}{\tb{POLYPHASE}} \\
    {}
      &&& \tb{More} 
      &&& \tb{More} \\
    \tb{(Constant)}
      & \tb{Slow} & \tb{Fast} & \tb{Elaborate} 
      & \tb{Slow} & \tb{Fast} & \tb{Elaborate} \\
    \hline\hline
    \mcol{7}{l}{\tb{ISLR (dB):} \cite{holm1991performance}} \\
    $-18.26$ 
      & $-16.11$ & $-16.08$ & $-16.07$
      & $-17.61$ & $-17.89$ & $-17.85$ \\ 
    \hline
    \mcol{7}{l}{\tb{ISLR (dB):} \cite{zenere2012sar}} \\
    $-20.51$ 
      & $-17.13$ & $-17.09$ & $-17.07$ 
      & $-20.43$ & $-20.43$ & $-20.44$ \\ 
    \hline
    \mcol{7}{l}{\tb{PSLR (dB):}} \\
    $-32.18$
      & $-29.02$ & $-28.91$ & $-28.93$
      & $-31.09$ & $-31.19$ & $-27.25$ \\
    \hline
    \end{tabular}
  \label{tab:MissingSamples}
\end{table}

\begin{figure}[htpb]
  \centering
  \subfigure[Different PRFs $\{8300, 8200\}$.]{%
    \includegraphics[width=0.45\linewidth]{%
      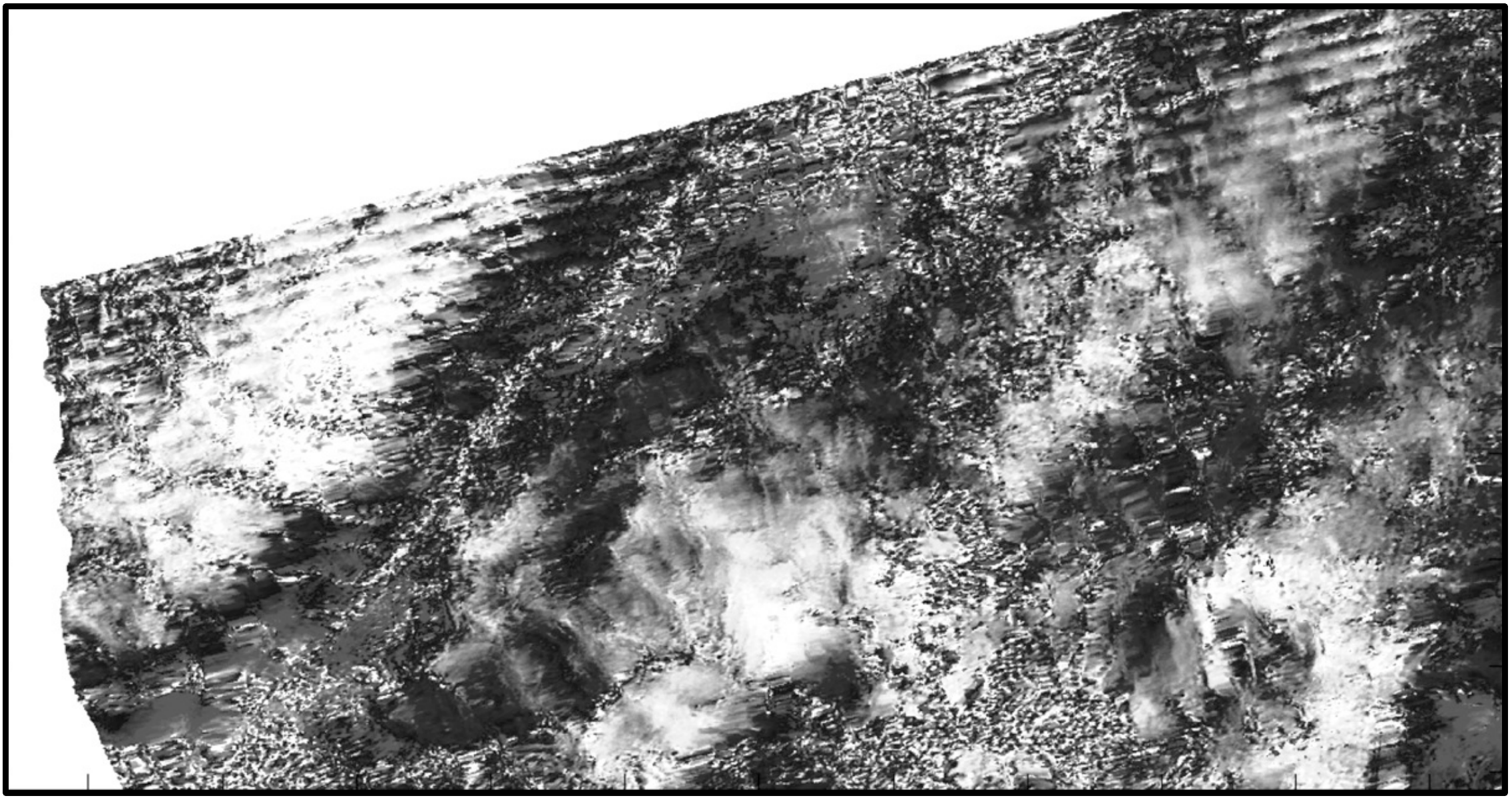}}
  \subfigure[Same PRFs $\{8300, 8300\}$.]{%
    \includegraphics[width=0.45\linewidth]{%
      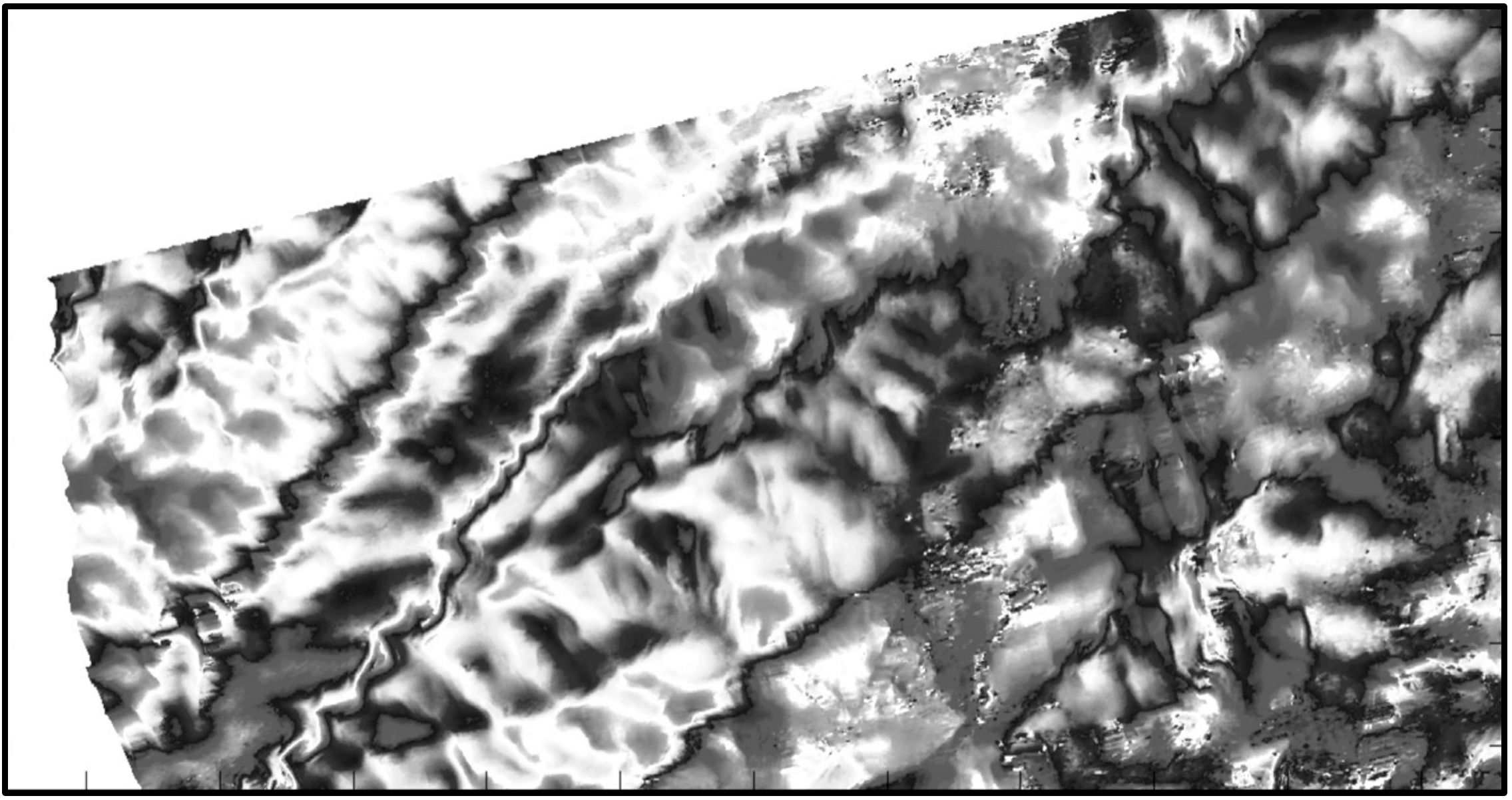}}
  \vv
  \caption{\scriptsize{Georeferenced interferograms of TerraSAR-X HS mode formed with SLCs. The phase has been wrapped to be within $[-\pi, +\pi]$.}}
  \label{Fig:TSXinterferograms}
\end{figure}

\bi{Other Benefits.} 
One potential use of POLYPHASE is to account for different PRFs across different acquisitions in repeat-pass Interferometric SAR (InSAR), a challenge that has been identified in the spotlight (SL), high resolution spotlight (HS), and the recent staring spotlight SAR modes of TerraSAR-X. These modes may not be able to collect data from the same scene at the same PRF \cite{Eineder2009ToGRS}, which poses a significant challenge in interferometry. POLYPHASE can be employed to unify the different PRFs across different SLCs, so that interferometry can be applied with equivalent spectrum widths. Fig.~\ref{Fig:TSXinterferograms}(a) shows the effect of this phenomenon on real data from TerraSAR-X spotlight interferometry; Fig.~\ref{Fig:TSXinterferograms}(b) shows the absence of this effect when the same PRF is used.

The popular solution to unify different SLCs to a common grid space and Doppler spectrum width is to undertake co-registration in two steps: 
  \tb{(a)}~resample the slave image in the image domain to the geometry of the master image information and low-resolution digital elevation model; 
  \tb{(b)}~estimate the residual shifts in range and azimuth within sub-pixel accuracy via point-like scatterers that are common to both images \cite{Adam2003ISPRS}. The success of this strategy depends on a high persistence of the scatterers in order to remove all residual shifts. The second step becomes more challenging at higher resolutions when gaining sub-pixel accuracy involves cells smaller than 1 (m). POLYPHASE resolves this issue directly without the need for such meticulous co-registration techniques. 
  
More importantly, POLYPHASE can in general be exploited in situations where the PRI sequence may not be periodic and the effects of non-uniform sampling across the aperture have to be compensated for, e.g., missing data, flight path deviation, imaging while in turn, and acceleration and deceleration. A case in point is dual-aperture SAR processing. Traditionally, coherent change detection (CCD) and ground moving target indicator (GMTI) algorithms, both of which are based on dual-aperture SAR processing algorithms, remove clutter by subtracting the SAR images formed within each aperture \cite{Soumekh1997ToIP}. If the sampling rates are not uniform across apertures, then the performance of clutter cancellation algorithms can be significantly diminished \cite{Skolnik1980}. POLYPHASE is an effective solution for compensating for this phenomenon and it does not assume any periodicity condition on the PRI sequence.


\section{Conclusion}
\label{sec:Conclusion}

POLYPHASE is a computationally efficient method for resampling along-track oversampled SAR data in slow-time domain for a radar that operates at variable PRFs. We provide a lower bound on the sparseness of the received SAR data relative to the output grid which ensures that the uniformly resampled data approximates the spectral properties of a decimated version of a certain ÔhiddenÕ densely sampled SAR data sequence. In essence, we view the non-uniformly spaced received samples as a subset of samples of a uniformly densely sampled underlying signal. A low-pass filter is then used to get the missing sample values. Only the portion of interest from the spectrum is extracted in the frequency domain after taking the spatial azimuth compression FFT. The filter implementation is carried out using its polyphase components. The order of each polyphase sub-filter and the polyphase implementation are critical factors affecting the computational complexity of the algorithm. 

When compared with BLUI in \cite{Villano2014ToGRS}, POLYPHASE provides significant savings in computational cost without sacrificing performance. It can be implemented in real-time and completely on-board with no down-linking of intermediate variables for on-ground computations. It can even accommodate out-of-order input samples. POLYPHASE can also be useful in other application scenarios, e.g., it can be employed to unify PRFs across a sequence of repeat-pass acquisitions taken at different PRFs in TerraSAR-X spotlight mode data, and it can also be used to improve clutter cancellation in coherent change detection and ground moving target indication.


\begin{appendices}


\section{Analytical Basis of the Resampling Scheme}
\label{appA}

As argued in Sections \ref{subsec:nonuniform} and \ref{subsec:3c}, we model the signal $r(n)$ in Fig.~\ref{Fig:UnifiedBlockDiagram} as $r(n)=g(n)\,s(n)$. Here, the gating function $g(n)$ is a realization of the i.i.d. Bernoulli random process with parameter $p$ in \eqref{eq:pBern}. So, $r(n)=s(n)$ whenever $g(n)=1$, and $r(n)=0$ (i.e., $r(n)$ is `missing' a sample of $s(n)$) otherwise. The mean of the w.s.s. random process $g(\cdot)$ is $\mu_g=p,\,\forall n\in\mbb{N}$; its autocorrelation $C_g(n)=p(1-p)\,\delta(n)+p^2$ and PSD $S_g(\omega)$ form a DT FT pair so that  
\begin{equation}
  S_g(\omega)
    =p(1-p)
       +(2\pi p^2)
        \sum_{k=-\infty}^{+\infty} \delta_D(\omega-2\pi k).
  \label{eq:Sg}
\end{equation}

With $f(n)$ being the IPR of the digital filter $F(z)$, we have
\begin{align}
  r(n)
    &=g(n)\,s(n);\;\;
  v(n)
    =f(n)\ast r(n);
      \notag \\
  v(n)
    &=(f(n)\ast g(n))\,v'(n).    
  \label{eq:rv}
\end{align}
Note that, $f(n)\leftrightarrow F(\omega)$. So, 
\begin{equation}
  y(n)
    =v'(nL)
    =\left.
       (f(n)\ast r(n))/(f(n)\ast g(n))
     \right|_{n\to nL}.
  \label{eq:y}
\end{equation}

In terms of PSDs, we can express \eqref{eq:rv} as
\begin{equation}
  S_r(\omega)
    =\frac{1}{2\pi}\,
     (S_g(\omega)\ast S_s(\omega));\;\;
  S_v(\omega)
    =|F(\omega)|^2S_r(\omega).
  \label{eq:SRV}
\end{equation}
The normalization step in \eqref{eq:rv} and \eqref{eq:y} can be expressed as 
\begin{align}
  S_v(\omega)
    &=\frac{1}{2\pi}\,
      ((|F(\omega)|^2S_g(\omega))\ast S_{v'}(\omega));
      \notag \\
  S_y(\omega)
    &=\frac{1}{L} 
      \sum_{\ell=0}^{L-1} S_{v'}(\omega_{\ell}),\;
      \omega_{\ell}
        =\frac{\omega}{L}-\frac{2\pi\ell}{L}.
  \label{eq:VY}
\end{align}
Compare the expressions for $S_v(\omega)$ in \eqref{eq:SRV} and \eqref{eq:VY}:
\begin{align}
  S_v(\omega)
    &=\frac{1}{2\pi}\,|F(\omega)|^2(S_g(\omega)\ast S_s(\omega))
      \notag \\
    &=\frac{1}{2\pi}\,(|F(\omega)|^2S_g(\omega))\ast S_{v'}(\omega).
  \label{eq:EQUAL}
\end{align}
Use \eqref{eq:Sg} to substitute for $S_g(\omega)$:
\begin{align}
  S_v(\omega)
    &=\frac{p(1-p)}{2\pi}\,|F(\omega)|^2
      \int_{\theta=-\pi}^{+\pi} S_s(\omega-\theta)\,d\theta
      \notag \\
    &\qquad
        +\frac{2\pi p^2}{2\pi}
         |F(\omega)|^2S_s(\omega)
      \label{eq:Sv_1} \\
    &=\frac{p(1-p)}{2\pi}
      \int_{\theta=-\pi}^{+\pi} |F(\theta)|^2S_{v'}(\omega-\theta)\,d\theta
      \notag \\
    &\qquad
        +\frac{2\pi p^2}{2\pi}\,|F(0)|^2 S_{v'}(\omega).
      \label{eq:Sv_2}
\end{align}

\begin{claim}
\label{cl:App} 
If $|F(\omega)|$ has support $[-\pi/L, +\pi/L],\,L>>1$, and \eqref{eq:specs} is true, then the PSD of the output $y(n)$ approximates the PSD of an $L$-fold decimated version of the densely sampled input signal $s(n)$ in $|\omega|\leq\gamma\,\pi$ when $p>>\rho/(L+\rho)$.
\hspace{\fill}
\IEEEQEDopen
\end{claim}

\IEEEproof
With $L>>1$, we note that
\begin{multline*}
  \int_{\theta=-\pi}^{+\pi} 
  |F(\theta)|^2S_{v'}(\omega-\theta)\,d\theta \\
    \leq
     \int_{\theta=-\pi/L}^{+\pi/L} 
     S_{v'}(\omega-\theta)\,d\theta 
    \approx 
     (2\pi/L)\,S_{v'}(\omega).
\end{multline*}
In \eqref{eq:Sv_2}, the first term is much smaller than the second term if $p>>1/(L+1)$. Then \eqref{eq:Sv_1} and \eqref{eq:Sv_2} become
\begin{multline*}
  S_v(\omega)
    =\frac{p(1-p)}{2\pi}\,|F(\omega)|^2
     \int_{\theta=-\pi}^{+\pi} S_s(\omega-\theta)\,d\theta \\
       +\frac{2\pi p^2}{2\pi}
        |F(\omega)|^2S_s(\omega)
    \approx
     \frac{2\pi p^2}{2\pi}\,S_{v'}(\omega).
\end{multline*}

Now, using the expression for $S_y(\omega)$ in \eqref{eq:VY}, consider 
\begin{align}
  \frac{1}{L} \sum_{\ell=0}^{L-1} S_v(\omega_{\ell})
    &=\frac{p(1-p)}{2\pi L}
      \sum_{\ell=0}^{L-1}
      |F(\omega_{\ell})|^2
      \int_{\theta=-\pi}^{+\pi} \!\!\!\!\!\! 
      S_s(\omega_{\ell}-\theta/L)\,d\theta
      \notag \\
    &\qquad
        +\frac{2\pi p^2}{2\pi L} \sum_{\ell=0}^{L-1}
         |F(\omega_{\ell})|^2S_s(\omega_{\ell})
      \label{eq:Sv_41} \\
    &\approx
      \frac{2\pi p^2}{2\pi L} \sum_{\ell=0}^{L-1} 
      S_{v'}(\omega_{\ell})
     =\frac{2\pi p^2}{2\pi}\,S_y(\omega).
      \label{eq:Sv_42}
\end{align}
Consider \eqref{eq:Sv_41}: for $|\omega|\leq\gamma\,\pi$, $|F(\omega_{\ell})|=1$. In addition, given that $S_s(\omega)$ has the support $[-\rho\pi/L, +\rho\pi/L]$, where $\rho=PRF_{\tx{in}}/PRF_{\tx{out}}\geq 1$ (see Fig.~\ref{Fig:Y_and_S}(b)), we may write  
\begin{equation}
  \int_{\theta=-\pi}^{+\pi} S_s(\omega_{\ell}-\theta/L)\,d\theta
    \approx
     (2\rho\pi/L)\,S_s(\omega_{\ell}),
\end{equation}
by integrating over $\theta\in[-\rho\pi/L, +\rho\pi/L]$ only. Thus, for $|\omega|\leq\gamma\,\pi$, we may approximate \eqref{eq:Sv_41} as
\[
  \frac{1}{L} \sum_{\ell=0}^{L-1} S_v(\omega_{\ell})
    \approx
     \frac{2\pi p^2}{2\pi L} \sum_{\ell=0}^{L-1}
        S_s(\omega_{\ell}),\tx{ for }
  p
    >>\rho/(L+\rho).
  \label{eq:Sv_51}
\]
Use this instead of \eqref{eq:Sv_41} to express \eqref{eq:Sv_41}-\eqref{eq:Sv_42} as
\[
  \frac{1}{L} \sum_{\ell=0}^{L-1} S_v(\omega_{\ell})
    \approx
     \frac{2\pi p^2}{2\pi L} \sum_{\ell=0}^{L-1}
     S_s(\omega_{\ell})
    \approx
     \frac{2\pi p^2}{2\pi}\,S_y(\omega).
  \tag*{\IEEEQEDclosed}
\]


\section{Operation of the Polyphase Components}
\label{appB}

\begin{claim}
\label{cl:xm}
Consider a non-zero input sample $r(\ell^*)\neq 0$ s.t. $\ell^*\in\mbb{N}$ and $(n-N_{pr})L\leq\ell^*\leq nL$. The only polyphase component that operates on $r(\ell^*)$ is $f_x(\cd)$, where $x=m^*L-\ell^*$, with 
\[
  m^*
    =\left\lfloor
       {\frac{\ell^*-1}{L}}
     \right\rfloor+1
  \implies
  x
    =(L-\ell^*)
       +L
        \left\lfloor
          {\frac{\ell^*-1}{L}}
        \right\rfloor.
  \tag*{\QEDopen}
\]
\end{claim}

\IEEEproof
First, suppose $(n-N_{pr})L+1\leq\ell^*\leq nL$ which corresponds to the summation term in \eqref{eq:exp2}. The polyphase components which operate on $r(\ell^*)$ are $f_x(\cd)$, where $x=m^*L-\ell^*\in\ol{0,L-1}$ with $m^*\in\ol{n-N_{pr}+1,n}$. But $m^*L-\ell^*=x$ iff $(m^*-1)L+(L-1-x)=\ell^*-1$. Since $(L-1-x)\in\ol{0,L-1}$, we conclude that $L-1-x=(\ell^*-1)_L$. This yields $x=(L-1)-(\ell^*-1)_L=(L-\ell^*)+L\left\lfloor{(\ell^*-1)/L}\right\rfloor$. The claim then follows for $(n-N_{pr})L+1\leq\ell^*\leq nL$.

Next, suppose $\ell^*=(n-N_{pr})L$ which corresponds to the first term in \eqref{eq:exp2}, viz., $f_0(N_{pr})\,r((n-N_{pr})L)$. When $\ell^*=(n-N_{pr})L$ is substituted in the claimed expressions for $x$ and $m^*$, we get $x=0$ and $m^*=n_{pr}$, which are consistent with $f_0(N_{pr})\,r((n-N_{pr})L)$. 
\hspace{\fill}\IEEEQEDclosed

\end{appendices}


\bibliographystyle{%
  ./bib/IEEEtran}
\bibliography{%
  ./bib/Torres2014ToGRS_references,%
  ./bib/IEEEabrv,%
  ./bib/articles_journals,%
  ./bib/articles_conferences,%
  ./bib/articles_books,%
  ./bib/articles_other,%
  ./bib/articles_urls}

  
\begin{IEEEbiography}{Yoangel Torres} 
(Student Member'04) received his B.Sc. degree in Electrical Engineering in (2005) from Florida State University, Tallahassee, Florida, and his M.S. in Electrical Engineering (2009) from the University of Florida, Gainesville, Florida. Presently, he is a Ph.D. student in the Department of Electrical and Computer Engineering at the University of Miami, Coral Gables, Florida. He is as a Radar Systems Engineer at Northrop Grumman Corporation (ES), Melbourne, Florida, which supports his Ph.D. studies. His research interests include, radar systems design, radar mode development, signal processing, and SAR/InSAR systems and their application in tunnel detection.
\end{IEEEbiography}

\begin{IEEEbiography}{Kamal Premaratne} 
(SM'94) received the B.Sc. degree in Electronics and Telecommunication Engineering (1982) with First-Class Honors from University of Moratuwa, Sri Lanka. He obtained his M.S. (1984) and Ph.D. (1988) degrees, both in Electrical and Computer Engineering, from the University of Miami, Coral Gables, Florida, where he is presently the Victor P. Clarke Professor.

He has received the ``Mather Premium'' (1992/93) and ``Heaviside Premium'' (1999/00) of the Institution of Electrical Engineers (IEE), London, UK, and the ``Eliahu I. Jury Excellence in Research Award'' (1991, 1994, 2001) and the ``Johnson A. Edosomwan Researcher of the Year Award'' (2014) of the College of Engineering, University of Miami. He has served as an Associate Editor of the {\sc IEEE Transactions on Signal Processing} (1994-1996) and the \emph{Journal of the Franklin Institute} (1993-2005). He is a Fellow of IET (formerly IEE). His current research interests include Dempster-Shafer (DS) belief theory, evidence fusion, machine learning and knowledge discovery from imperfect data, and opinion and consensus dynamics in social networks.
\end{IEEEbiography}

\begin{IEEEbiography}{Falk Amelung} 
received the M.S. degree in Geophysics from the University of M\"unster, M\"unster, Germany, in 1992 and the Ph.D. degree from the University of Strasbourg, Strasbourg, France, in 1996. After several years as a Research Scientist with Stanford University, Stanford, California, and the University of Hawaii, Honolulu, he joined the University of Miami, Miami, Flotida, in 2002, where he is currently a Professor of Geophysics in the Department of Marine Geosciences, Rosenstiel School of Marine and Atmospheric Science. His research expertise lies in InSAR study of active volcanism, active tectonics, and glacial rebound land subsidence. His particular research goal is the use of InSAR for routine volcano monitoring. 
\end{IEEEbiography}

\begin{IEEEbiography}{Shimon Wdowinski} 
is an Associate Professor at the Department of Earth \& Environment, Florida International University, where he teaches and researches geology and geophysics. His work has focused on the development and usage of space geodetic techniques that can detect very precisely small movements of the EarthÕs surface. He successfully applied these technologies to study natural hazards and environmental phenomena, such as earthquakes, land subsidence, and wetland surface flow. He received a B.Sc. in Earth Sciences (1983) and M.Sc. in Geology (1985) from the Hebrew University (Jerusalem, Israel) and an M.S. in Engineering Sciences (1987) and Ph.D. in Geophysics (1990) from Harvard University. He conducted a post-doctorate research at Scripps Institute of Oceanography, UCSD (1990-1993), served as Associate Professor at Tel Aviv University (1994-2004), served as a Research Professor at the Rosenstiel School of Marine and Atmospheric Sciences, University of Miami (2004-2016), and joined Florida International University in 2016.
\end{IEEEbiography}
\vfill

\end{document}